\documentclass[]{aa}  
\usepackage{graphicx}
\usepackage{txfonts}
\usepackage{natbib}
\bibpunct{(}{)}{;}{a}{}{,} % to follow the A&A style
\def\lsx{$L_{\rm{soft X}}$}
\def\lx{$L_{\rm{X}}$}
\def\lfir{$L_{\rm{FIR}}$}
\def\xeff{$\epsilon_{\rm{xeff}}$}
\def\emech{{$\dot{E}_{K}$}}
\def\lradio{$L_{ \rm{rad}}$}
\def\lnth{$L_{\rm{nth}}$}
\def\lth{$L_{\rm{th}}$}
\def\sfr{{\em SFR}}
\def\sfs{{\em SFS}}
\def\ergs{erg s$^{-1}$}

\def\msun{M$_\odot$}
\def\ebv{{\em E(B-V)}}
\def\nlyc{{$N_{\rm{Lyc}}$}}
\def\lyalp{{Ly$\alpha$}}
\def\lulyalp{{$L(\rm{Ly}\alpha)$}}

\def\luv{{$L_{UV}$}}
\def\lcuva{{$L_{1500}$}}
\def\lcuvb{{$L_{2000}$}}
\def\lgu{{$L_{3500}$}}
\def\lgb{{$L_{4400}$}}
\def\lgv{{$L_{5500}$}}
\def\lgk{{$L_{22200}$}}
\def\ewwr{{$EW($WRbump$)$}}
\def\ewhb{{$EW(\rm{H}\beta)$}}
\def\sfrfir{{\em SFR\rm{(FIR)}}}

\def\sfrnlyc{{\em SFR$(N_{\rm{Lyc}})$}}
\def\sfrha{{\em SFR$(\rm{H}\alpha)$}}
\def\sfrrad{{\em SFR($L_{ \rm{rad}}$)}}
\def\alphanth{$\alpha_{\rm{nth}}$}
\def\alphath{$\alpha_{\rm{th}}$}

\def\msfr{M$_\odot$ yr$^{-1}$}
\def\hal{{$\rm{H}\alpha$}}
\def\hb{{$\rm{H}\beta$}}
\def\lha{{$L(\rm{H}\alpha)$}}
\def\lhb{{$L(\rm{H}\beta)$}}

\begin{document}
\title{Calibration of star formation rate tracers for short- and long-lived star formation episodes}

\titlerunning{Calibration of star formation rate tracers}

   \author{H. Ot\'{\i}-Floranes\inst{1,2}
          \and
          J.M. Mas-Hesse\inst{1}
}

   \offprints{J.M. Mas-Hesse}

   \institute{Centro de Astrobiolog\'{\i}a -- LAEX (CSIC--INTA), POB 78, 
	     28691 Villanueva de la Ca\~nada, Spain\\
             \email{otih@cab.inta-csic.es, mm@cab.inta-csic.es}
\and
	     Dpto. de F\'{\i}sica Moderna, Facultad de Ciencias, Universidad de Cantabria, 39005
	      Santander, Spain
             }

   \authorrunning{Ot\'{\i}-Floranes \& Mas-Hesse}

   \date{Received; accepted}
 
  \abstract
%context heading (optional)
{To derive the history of star formation in the Universe a set of calibrated star formation rate tracers at different wavelengths is required. The calibration has to consistently take into account the effects of extinction, star formation regime (short or long-lived) and the evolutionary state to avoid biases at different redshift ranges. }
{We use evolutionary synthesis models optimized for intense episodes of star formation to compute a consistent calibration of the most usual star formation rate tracers at different energy ranges, from X-ray to radio luminosities. } 
{We have computed the predicted evolution of the different estimators taking into account nearly-instantaneous and continuous star formation regimes and the effect of interstellar extinction (attenuation at high energies, thermal reradiation in the far infrared). We have also considered the effect of metallicity on the calibration of the different estimators.  }
{A consistent calibration of a complete set of star formation rate tracers is presented, computed for the most usual star-forming regions conditions in terms of evolutionary state, star formation regime, interstellar extinction and initial mass function. We discuss the validity of the different tracers in different star formation scenarios and compare our predictions with previous calibrations of general use. }  
{In order to measure the intensity of star formation episodes we should distinguish between nearly-instantaneous and continuous star formation regimes. While the {\em star formation strength} \rm (\msun) should be used for the former, the more common {\em star formation rate} \rm (\msun\ yr$^{-1}$)  is only valid for episodes forming stars at a constant rate during extended periods of time. Moreover, even for extended star formation episodes, the evolutionary state should be taken into account, since most \sfr\ tracers stabilize only after 100 Myr of evolution.  }

\keywords{ Stars: formation -- Galaxies: evolution --  Galaxies: fundamental parameters -- Galaxies: luminosity function, mass function --  Galaxies: starburst -- Galaxies: stellar content }

   \maketitle

\section{Introduction} 

The availability of a continuously increasing observational data set on galaxies at higher and higher redshifts allows the study of the history of star formation at cosmic scales, approaching already (presently up to $z \ge 7$)  the ages of reionization when the first population of massive stars started to ignite. The star formation rate (\sfr) of these galaxies at different ages is derived from the so-called star formation rate tracers, spatially integrated parameters at different wavelengths that allow the determine of the strength of the star formation episode by comparison with the predictions of evolutionary synthesis models. \citet{Kennicutt98} discussed the uncertainties inherent to the use of optical broad-band colors, and presented an updated calibration of the UV continuum, recombination lines (or number of ionizing photons emitted per unit time), forbidden lines and total far infrared emission that has become {\em de facto} the reference standard calibration of these estimators. 

While widely used, it is often neglected that the \sfr\ tracers calibration proposed by \citet{Kennicutt98} was computed for a very specific star formation history, i.e., a star formation episode running at a nearly constant rate during a long enough period of time ($>100$ Myr), so that an equilibrium is reached between the number of massive stars dying and igniting. Under these conditions most of the parameters considered reach an equilibrium state with almost no evolution with time. This {\em continuous burst} scenario is certainly valid for many galaxies, especially large spirals in which massive star formation takes place at  {\em spatially integrated} nearly constant rates during long periods of time (though at different locations within their spiral arms). But star formation seems to proceed in a more {\em bursty} regime in other galaxies, showing extremely high present-day star formation rates, indicating that they might be experiencing a nearly {\em instantaneous} (i.e., extended over few million years) massive star formation episode.  Under these circumstances, the derivation of star formation rates by comparison with the predictions by continuous bursts might be erroneous. Moreover, even the concept of a star formation rate, measured as \msfr, might be misleading in these cases. The intensity of these massive star formation episodes should be better parameterized by the {\em star formation strength} (\sfs), measured directly in units of \msun, indicating the total amount of mass having been transformed into stars. 

On the other hand, as noted by \citet{Rosa02}, different studies have indicated the significant role played by dust in the estimates of star formation rates. Extinction by dust significantly alters the integrated multiwavelength spectrum escaping from a star-forming region, weakening the far-UV continuum, but also boosting the emission in the far infrared range. It is critical therefore to take into account the effects of extinction by dust in a consistent way, since the calibration of a multiwavelength set of \sfr\ and \sfs\ tracers will be a strong function of the dust abundance.   

In this paper we present an updated calibration of a complete set of star formation rate tracers at different wavelengths, from X-rays to radio, computed in a consistent way with state-of-the-art evolutionary synthesis models considering different star formation histories, and taking into account the effects of dust extinction at all wavelengths. In Sect.~2 we summarize the properties of the tracers considered in this work, and in Sect.~3 we describe the evolutionary synthesis models we have used. In Sect.~4 we present the calibration of the different estimators, discuss their sensitivity to the properties of the star formation scenario and compare our predictions with other evolutionary codes. Finally,  Sect.~5 summarizes our main conclusions.

\section{Star formation rate tracers}

We have selected a large sample of \sfr\ and \sfs\ tracers covering a broad wavelength range, from
X-rays to radio, and associated to different physical processes directly linked to the strength of
the star formation episode. We want to remark that most of these tracers may be contaminated
by emission not related to the present episode of star formation, but associated to an existing AGN,
underlying older stellar populations, evolved but hot, low-mass stars, etc. During this work we will
consider that the different star formation rate tracers have been corrected by the user before
applying the calibrations. Since these corrections will not always be accurate or even possible, the
\sfr\ values so derived  have to be taken with the corresponding caution. The reliability of the estimates will be larger the more \sfr\ tracers can be used simultaneously for a given object, since the  contaminating effects have different strengths at different energy ranges.

\subsection{Far infrared luminosity  (\lfir)}

Most radiation emitted by the young, massive stars is absorbed by the dust particles surrounding the star formation region, principally in the UV. The heated dust reemits this energy in the far infrared (FIR) range (which we will consider as $1-1000 ~\mu\rm{m}$), especially within the  $10-120 ~\mu\rm{m}$ domain. The fraction of energy emitted by these stars in the UV/optical which is absorbed by dust and reemitted as FIR radiation is close to 1 for values of \ebv\ above 0.5 \citep{MasHesse91}. Hence, the FIR luminosity is in general a good estimator of the bolometric luminosity of the massive stars, and thus \sfrfir\ is one of the most used and reliable \sfr\ estimators. \citet{Kennicutt98} derived a calibration of \lfir\ using population synthesis models of extended bursts in their asymptotic phase (when the rate of massive star formation and death is balanced), assuming the optically thick case, but stating that the expression strictly applies to bursts younger than 100 Myr. This calibration would be altered if corrections for the dust heating from old stars in quiescent galaxies were made.  

Besides the total FIR luminosity, some other calibrations make use of the SED (spectral energy distribution)  of the burst at certain FIR wavelengths. \citet{AlonHer06} derived 
an empirical, non-linear calibration for $SFR(L(24 ~\mu\rm{m}))$ by combining $L(24 ~\mu\rm{m})$ observations of Luminous InfraRed Galaxies (LIRGs), Ultraluminous InfraRed Galaxies (ULIRGs), normal galaxies and HII regions within M51, and  the  $SFR(P\alpha)$ values derived from the \sfrha\ calibration from \citet{Kennicutt98} (assuming case B recombination).  \citet{Calzetti2007} found a similar relation based on luminosity surface density values of HII knots from 33 nearby galaxies.

\lfir\ can become contaminated by the presence of an AGN, especially when dealing with spatially-integrated luminosities of distant, unresolved galaxies. Depending on the redshift of the galaxy, the width of the Balmer or Lyman emission lines should be checked to exclude AGN dominated objects. In galaxies dominated by massive star-forming episodes, the \sfr\ values derived from \lfir\ and from \luv, computed with the same \ebv\ levels, should be consistent.

\begin{figure}
\centering
\includegraphics[width=8cm,bb=28 40 709 566
dpi,clip=true]{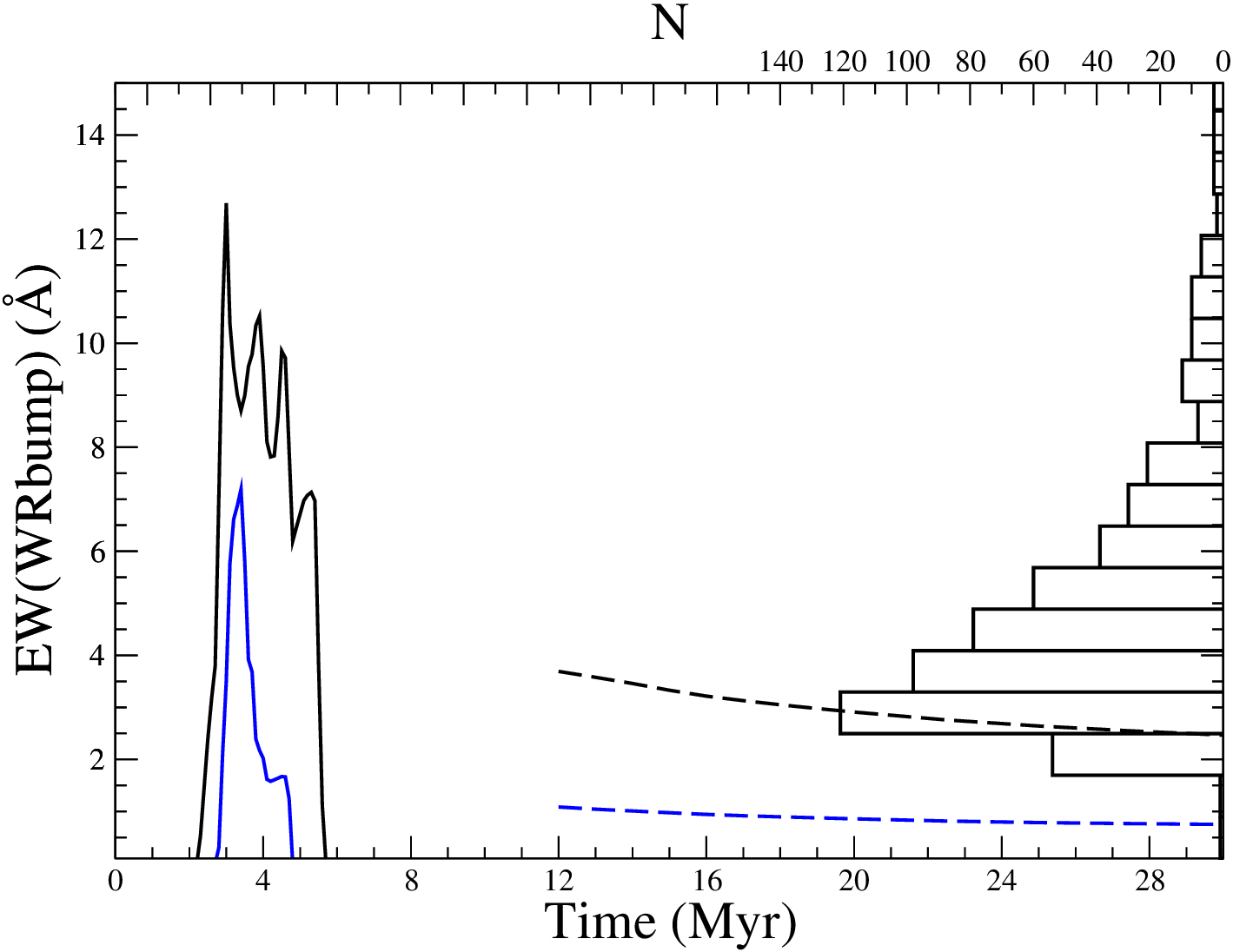}
\caption{Evolution of \ewwr\ predicted by CMHK02 for IB (solid lines) and EB (dashed lines) models both for $Z=0.020$ (top) and $Z=0.008$ (bottom), together with the histogram of the \ewwr\ data of the sample from \citet{Brinchmann08}.} 
\label{figEWWR}
\end{figure}

\subsection{Ionizing power (\nlyc) and emission lines}

Massive stars are conspicuous sources of ionizing photons, and therefore their number \nlyc\ could trace the burst intensity directly. However, practically all ionizing photons are absorbed by gas and dust within the nebular region. Recombination of the hydrogen atoms ionized by this radiation and free electrons produces intense emission lines, such as H$\alpha$ or H$\beta$, commonly used as \sfr\ estimators since their luminosity is proportional to \nlyc\ under basic assumptions \citep{Osterbrock89}. \citet{Kennicutt98} provided an expression for \sfrnlyc\ based on the calibrations by \citet{Kennicutt94} and \citet{Madau98}, and used it to derive a calibration of \sfrha\ assuming case B recombination and $T_e=10^4$ K. A somewhat different expression was derived by \citet{Rosa02}. They studied a sample of HII, starburst and blue compact galaxies, in order to obtain empirical \sfr\ calibrations based on \lha, $L(\rm{[OII]}3727)$ and $L_{UV}$.  \citet{Rosa02} obtained an \sfrha\ coefficient $40$\% lower than the value from \citet{Kennicutt98} since it was based on the observed \hal\ luminosity, not corrected for Balmer absorption due to the stellar population. 

H$\alpha$ has the disadvantage of falling outside the optical spectrum when studying medium-redshift sources ($z\geq0.4$). Studying a sample of $412$ local star-forming galaxies, \citet{Moustakas06} showed that, although weaker and more affected by extinction, H$\beta$ line can also be used to estimate \sfr. Applying the \lha\ calibration by \citet{Kennicutt98}, they calculated the \sfr\ values using both \lha\ and  \lhb,  assuming case B recombination (i.e. $L(\rm{H}\alpha)/L(\rm{H}\beta)=2.86$). They  obtained very similar results, with a $40$\% scatter.

The luminosity of the \lyalp\ line can also be used as a star formation rate estimator, since it is in principle a linear function of \nlyc, with the advantage that it is visible in the optical range for galaxies at a redshift of $z > 2.0$. Nevertheless, \lyalp\ photons are affected by resonant scattering in neutral Hydrogen, strongly affecting the intensity of the line, as we will discuss later.

\subsection{Continuum luminosity}

Most radiation from young, massive stars is emitted in the UV range (912 to around 3000 \AA) and could be used as a reliable \sfr\ and \sfs\ estimator since the contamination from older stellar populations is low or even negligible below 1800 \AA. However, attenuation correction plays an important role here, since UV photons are severly affected by extinction. \citet{Kennicutt98} uses the calibration from \citet{Madau98} in order to obtain an expression for \sfr\ as a function of $L_{\nu}$, assuming an extended star formation process with an age larger than $100$ Myr and solar metallicity. According to \citet{Kennicutt98}, the expression can be used in the wide range of $1500-2500$ \AA, where the spectrum of a stellar population with a Salpeter initial mass function (IMF) is rather flat (as a function of frequency). 

Fitting {\em GALEX} and SDSS (Sloan Digital Sky Survey) data of nearly $50000$ galaxies with SEDs constructed with the population synthesis models of \citet{Bruzual03}, \citet{Salim07} derived an expression for $SFR(\rm{UV})$ which yields \sfr\ values lower by $30$\% than those obtained with the  \citet{Kennicutt98} calibration. They claim that the reasons for this difference are the low metallicity of their sample ($Z=0.016$), the different star formation histories of the objects of the sample and some intrinsic differences between \citet{Bruzual03} and \citet{Madau98} models. 

Some authors \citep{Buat99, Hira03, IP04} have opted for merging in a single expression the observed values of UV and FIR luminosities. The latter component would represent the main bulge emission of massive stars, while the former would correct for the radiation which eventually escapes from the burst before it is absorbed by dust. This way, prior assumptions about extinction would not be necessary, but both the FIR and the UV luminosities would be needed, so that the method is not easily applicable for high-redshift galaxies.  

\citet{Kennicutt98} concluded that broad-band luminosities in the optical are poor \sfr\ estimators, since the optical continuum is contributed by stars at very different evolutionary states and affected by potentially different degrees of extinction. This is especially true for large spiral galaxies with a complex history of star formation. Even the UV continuum might be contaminated by blue horizontal-branch stars in those objects dominated by an old underlying stellar population, specially in the case of early-type galaxies. Nevertheless, continuum luminosities in the optical--near infrared bands could be reliable estimators for compact objects whose continuum is dominated by the present burst of star formation. We have calibrated \sfr\ and \sfs\ with the luminosities at 1500 \AA, 2000 \AA, 3500 \AA\ (U), 4400 \AA\ (B), 5500  \AA\ (V) and 2.2 $\mu$m (K).

\subsection{X-ray luminosity (\lx)}

Mechanical energy released by the  massive stars' stellar winds  and supernova explosions  heats the surrounding gas, originating a diffuse X-ray emission which peaks in the soft X-ray range. This diffuse emission adds to the X-ray point source radiation from X-ray binaries and supernova remnants (SNR). Whereas the emission from X-ray binaries dominates the hard X-rays range ($2-10$ keV), emission from diffuse hot gas drives the radiation in the soft range ($0.2-2$ keV) \citep{Cervino02}. Many authors have claimed recently that the X-ray emission from star-forming regions should be a direct function of the burst intensity, and several \sfr\ estimators based on different components and energy ranges of this radiation have been published. Analyzing the total X-ray emission from local starburst galaxies extracted from the \citet{Ho97} atlas, which had been observed by ASCA and/or BeppoSAX, \citet{Ranalli03} found both soft and hard X-rays \sfr\ linear expressions using FIR as a proxy, based on the \sfrfir\ relation by \citet{Kennicutt98}. A similar study was performed by \citet{Tullmann06b} using XMM-Newton and Chandra data, but in this case obvious point sources were removed. On the other hand, \citet{Grimm03} used the hard emission from  X-ray binaries of Chandra-resolved, nearby late-type/starburst galaxies, to derive a relation which becomes non-linear at low \sfr\ values. \citet{Rosa09} have recently analyzed XMM-Newton observations of a sample of 14 star-forming galaxies with high \sfr\ (0.2 -- 160 \msfr), as derived from radio, FIR and UV tracers, confirming that the \sfr\ derived from soft X-rays is comparable to that obtained from \hal\ luminosities. 

Using evolutionary synthesis models, \citet{MasHesse08} proved the importance of the evolutionary state of the burst when trying to ascertain the \sfr\ value via the soft X-ray luminosity, since it can increase between half and one order of magnitude within $20$ Myr in the models with constant \sfr. They derived two expressions, both for young  ($\sim10$ Myr) and more evolved extended star formation bursts ($\sim30$ Myr). They also provided a star formation strength expression for nearly instantaneous star-forming episodes, yielding the initial, total stellar mass of gas transformed into stars.  We have included the calibration of \lx\ in this paper for completeness. We refer to \citet{MasHesse08} for a deeper discussion.

\subsection{Radio luminosity (\lradio)}

A very tight correlation between \lfir\ and radio emission has been found over the past decades, common to a great variety of objects, such as starburst galaxies, normal spiral galaxies, blue compact dwarfs (BCDs), E/S0 galaxies, irregular galaxies, HII regions, etc... \citep{MasHesse92,Condon92,Bell03}, which proves that radio luminosity might be a reliable star formation rate estimator. Radio luminosity in star-forming regions is composed of both a thermal (\lth) and a non-thermal (\lnth) component. The former appears as a byproduct of free-free interactions, i.e. Bremmstrahlung, and free-bound transitions between the constituents of the ionized gas and its value can be expressed in terms of \nlyc\ \citep{Rubin68}, while the non-thermal emission is synchrotron in nature, emitted by electrons after being accelerated by supernovae explosions, and it is usually assumed to be proportional to the rate at which SN explode \citep{Ulvestad82}. Both contributions have different spectral index values, which are \alphath\ $=-0.1$ for \lth\ \citep{Rubin68} and \alphanth\ $=-(0.8 - 0.9)$ for \lnth\ (see \citet{MasHesse92} and references therein). $SFR$ calibrations based on both \lnth\ and \lth\ can be found in \citet{Condon92}, assuming $\alpha=-0.8$ for the former \citep{Condon90} and an electronic temperature $T_{e}=10^{4}$ K and no dust absorption for the latter. Applying the calibration of \sfrfir\ from \citet{Kennicutt98} to a sample with objects of very diverse natures (spiral galaxies, starburst galaxies, BCDs, irregular galaxies, etc), and after considering the tight correlation between infrared and radio emissions, \citet{Bell03} obtained a non-linear expression for \sfrrad\ using \lfir\ as a proxy.

\section{Evolutionary synthesis models}

In order to calibrate the different \sfr\ tracers in a consistent way, we have computed their expected values under different scenarios using the evolutionary population synthesis models of \citet{Cervino02} (hereafter CMHK02 models\footnote{Downloadable from {\rm http://www.laeff.inta.es/users/mcs/SED/}}),  \citet{Leitherer99} (Starburst99, hereafter SB99 models\footnote{Downloadable from {\rm http://www.stsci.edu/science/starburst99/}}) and \citet{Schaerer2002,Schaerer03} (SC02 models\footnote{Downloadable from {\rm http://obswww.unige.ch/sfr/sfr\_tls/pop32/ }}).

CMHK02 models, which are based on the models by \citet{Arnault89}, \citet{MasHesse91} and \citet{Cervino94}, compute the evolution of a young population of massive stars which are formed at the same time (instantaneous bursts, IB, also referred to in the literature as single stellar populations -- SSP) or which form at a constant rate during an extended period of time (extended bursts, EB). The different observables are calculated for the first $30$ Myr after the onset of the burst. They are scaled to the mass of gas transformed into stars at the start of the burst in IB models (measured in \msun), and to the mass of gas transfomed into stars per unit time for EB models (\msfr). The initial mass function (IMF)  of the stellar population is defined by a power-law within the mass range of 2--120 \msun\, with the slopes $\alpha=1\mbox{, }2.35\mbox{, }3$. The models include different metallicities,    $Z/Z_{\sun}=0.05\mbox{, }0.2\mbox{, }0.4\mbox{, }1\mbox{, }2$. For this work we have considered a nominal scenario with $\phi(m)\sim m^{-2.35}$ for the IMF (i.e. a Salpeter IMF), and solar metallicity $Z = Z_{\sun}$. The  presence of binary systems has not been considered. Chemical evolution is not treated self-consistently, since all stellar generations in EB models have the same metallicity. This should not be a problem, since the effect becomes important at large ages ($\sim1$ Gyr), as shown by \citet{Fioc97}. 

SB99 synthesis models \citep{Leitherer99} are based on the original models by \citet{Leitherer95}, but have been continuously updated with new evolutionary tracks and stellar atmospheres. Several evolutionary tracks with different metallicity values can be used, and for consistency with CMHK02 models, we opted for the Geneva standard mass-loss tracks. SB99 allows us to compute the observables up to 1 Gyr after the onset of the burst as well as the use of a step-IMF and different mass limits. As in CMHK02 models, both EB and IB star formation regimes are considered, and metallicity-evolution is not taken into account. We have used SB99 to compute the predicted values of the different parameters for EB episodes that have formed stars during more than 200~Myr at a constant rate, since the models by  CMHK02 are available only for the first 30 Myr. We have checked that where they overlap in time, the predictions by both sets of models are perfectly consistent. For the IB star formation regime we have preferred to use CMHK02 models, since they are based on Monte Carlo simulations of the IMF and consider therefore in a more realistic way the  
stochastical nature of the massive stellar populations. Moreover, since they provide directly the expected X-ray luminosity, as discussed below, the other \sfr\ tracers should be computed with the same models for consistency. 

Finally, SC02 models have been used for completeness to compute \luv\ and \nlyc\ for population III (Pop. III) stellar populations, which could correspond to the scenario valid for star-forming galaxies at very high redshift, where metallicity is extremely low ($Z\sim0$).

While most \sfr\ tracers calibrations are based on the predictions by evolutionary models which are computed for extended star formation processes where gas is transformed into stars at a nearly constant rate during long periods of time (hundreds to thousands Myr), we want to stress that star formation seems to proceed in the form of short, nearly instantaneous bursts in many galaxies. It is generally difficult to distinguish a short, but young starburst from an evolved, extended star formation process, since most observable parameters are related to just the most massive and therefore young stars in the region. Nevertheless, a long-lasting star formation process produces an accumulation of medium-low mass stars with long lifetimes,  which can dominate the optical continuum. As an effect, these kind of episodes show systematically lower equivalent width values of \hb\ and Wolf-Rayet star bumps at around 4686~\AA\ (hereafter \ewwr) than young instantaneous bursts \citep{Cervino94}. 

Figure \ref{figEWWR} shows that continuous star formation models in the asymptotic phase cannot reproduce the high wing of the distribution of \ewwr\ values measured by \citet{Brinchmann08} on their sample of 570 star-forming galaxies. Instantaneous bursts, on the other hand, of course predict the high \ewwr\ range (see also \citet{MasHesse99b}). The sample by \citet{Brinchmann08} is clearly biased towards star-forming galaxies with large Wolf-Rayet bump equivalent widths, i.e., experiencing short star formation episodes, but other authors have found evidences of short-lived bursts. \citet{Pellerin07}, for example, find that EB models can not reproduce the FUSE far-UV data of the $24$ starburst galaxies of their sample, whereas there is a much better agreement with the predictions by IB models.  

We conclude therefore that star formation proceeds in the form of nearly instantaneous, short bursts in a significant fraction of objects. In the case of very large, non-resolved galaxies, the superposition of individually instantaneous bursts at different evolutionary states can mimic the properties of an extended star formation process at a nearly constant rate, while large objects at high redshift could sustain intense star formation processes during truly extended periods of time. When deriving the intensity of the different star formation episodes in the history of a galaxy, its star formation regime has to be carefully taken into account. In the case of a nearly instantaneous burst, the ionizing power decreases very fast with time, and becomes negligible after the first 8-10 Myr \citep{MasHesse91}. Blindly applying an \sfr\ estimator related to \nlyc, calibrated for an extended star formation regime, can lead to discrepancies of several orders of magnitude depending on the evolutionary state. Moreover, even the concept of a star formation rate as such becomes completely misleading in these cases, since formation of massive stars might have ended already a few million years ago. The concept of a star formation strength, \sfs, measured as the total mass of gas transformed into stars during the burst, is much better adapted to this scenario. Unfortunately, deriving the \sfs\ by comparing the observations with the predictions of synthesis models requires a preliminary estimate of the evolutionary state of the burst, and this can be obtained only when parameters like the equivalent width of \hb\ can be measured.  We have therefore calibrated our set of \sfr\ and \sfs\ tracers for both star formation regimes, extended and instantaneous respectively, and for different evolutionary states of the burst. 

Formation of massive stars is an inherently stochastic process, so that when a relatively low number of stars are formed, we can not guarantee that the whole IMF will be filled. \citet{Cervino02a} showed that stochastic effects become small enough only for starbursts that have transfomed more than $10^{5-6}$~\msun\ of gas into stars.  The star formation estimators should therefore not be applied to smaller star-forming regions, where the stochastic effects could significantly affect the integrated emission.

In the sections below we describe the way the different estimators have been computed. More details can be found in \citet{MasHesse91}, \citet{Cervino94} and \citet{Cervino02}.

\begin{figure*}
\begin{center}
\includegraphics[width=8cm,bb=5 40 714 529
dpi,clip=true]{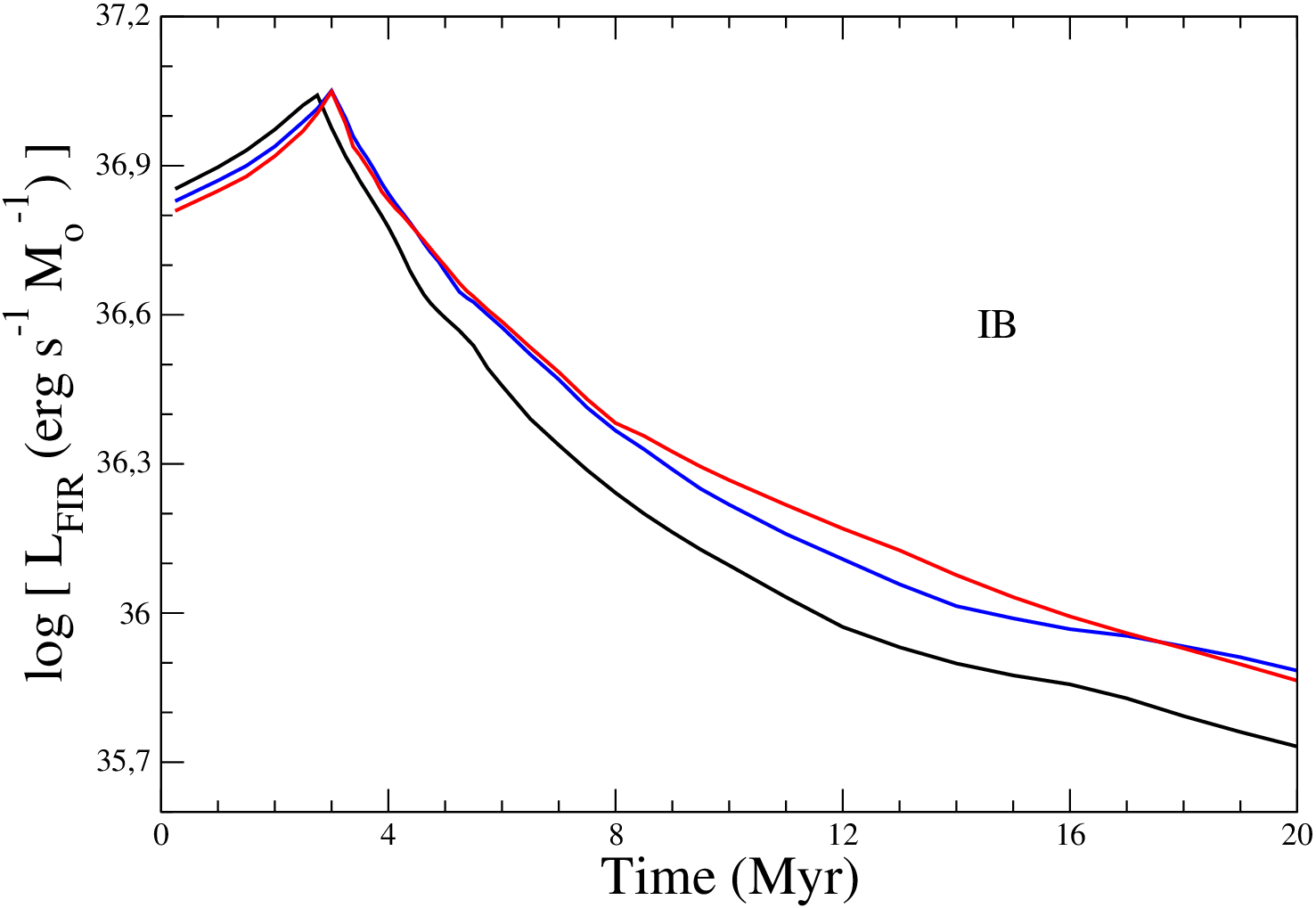}
\hspace{1.0 cm}
\includegraphics[width=8cm,bb=5 40 718 522
dpi,clip=true]{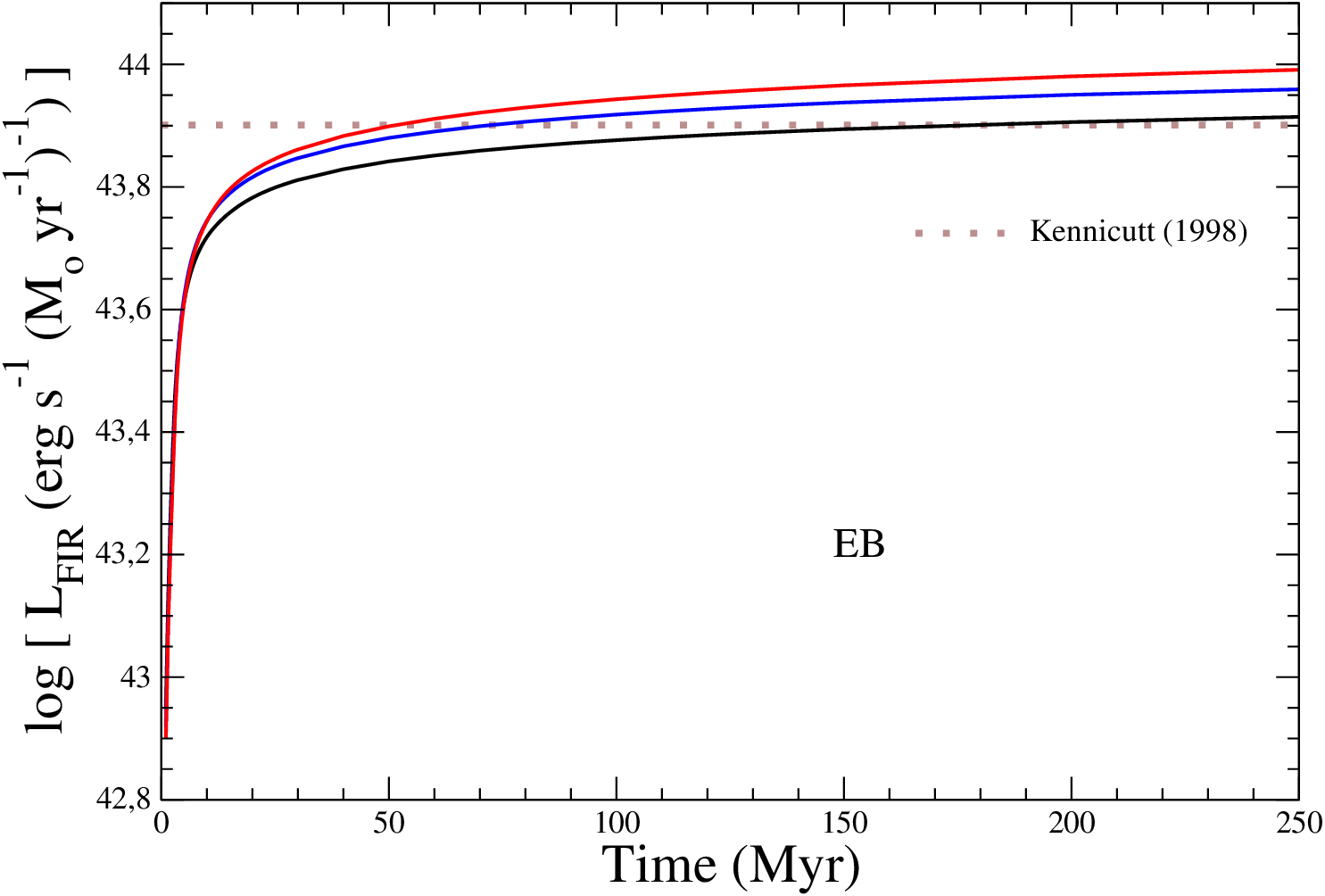}
\end{center}
\caption{Evolution of \lfir\ predicted by CMHK02 for instantaneous (IB) models (left, normalized to \sfs\ $= 1$ \msun) and by SB99 for extended (EB) models (right, normalized to \sfr\ $= 1$ \msun\ yr$^{-1}$). \lfir\ has been computed for \ebv$=1.0$ and three metallicity values (from bottom to top:  $Z=0.020\mbox{-black, }0.008\mbox{-blue, }0.001\mbox{-red}$). We also plot on the right panel the \lfir\ calibration by \citet{Kennicutt98}, adapted to the IMF used in this work.}
\label{figlfir}
\end{figure*}

\begin{figure*}
\begin{center}
\includegraphics[width=8cm,bb=7 40 708 522
dpi,clip=true]{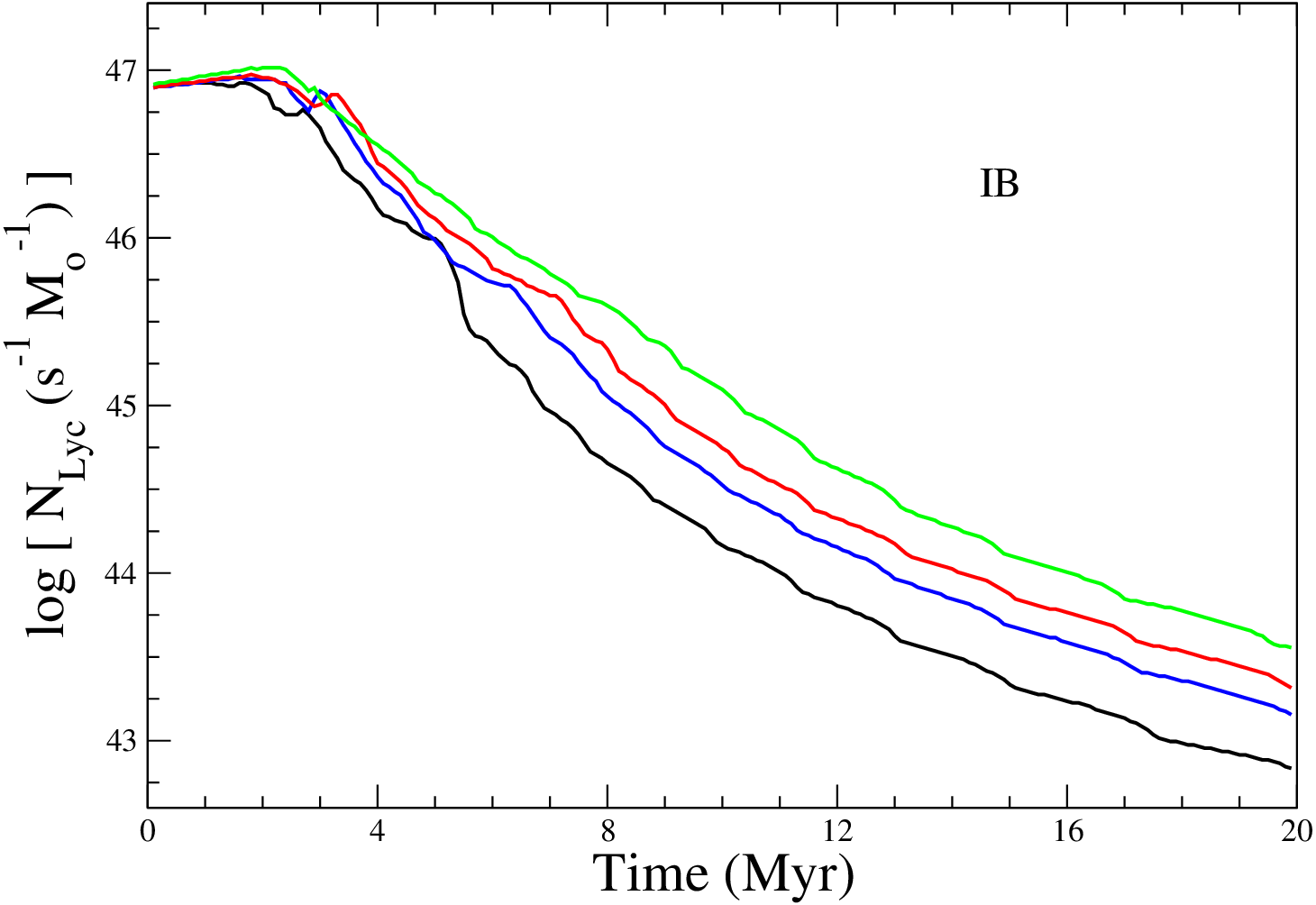}
\hspace{1.0 cm}
\includegraphics[width=8cm,bb=1 40 708 522
dpi,clip=true]{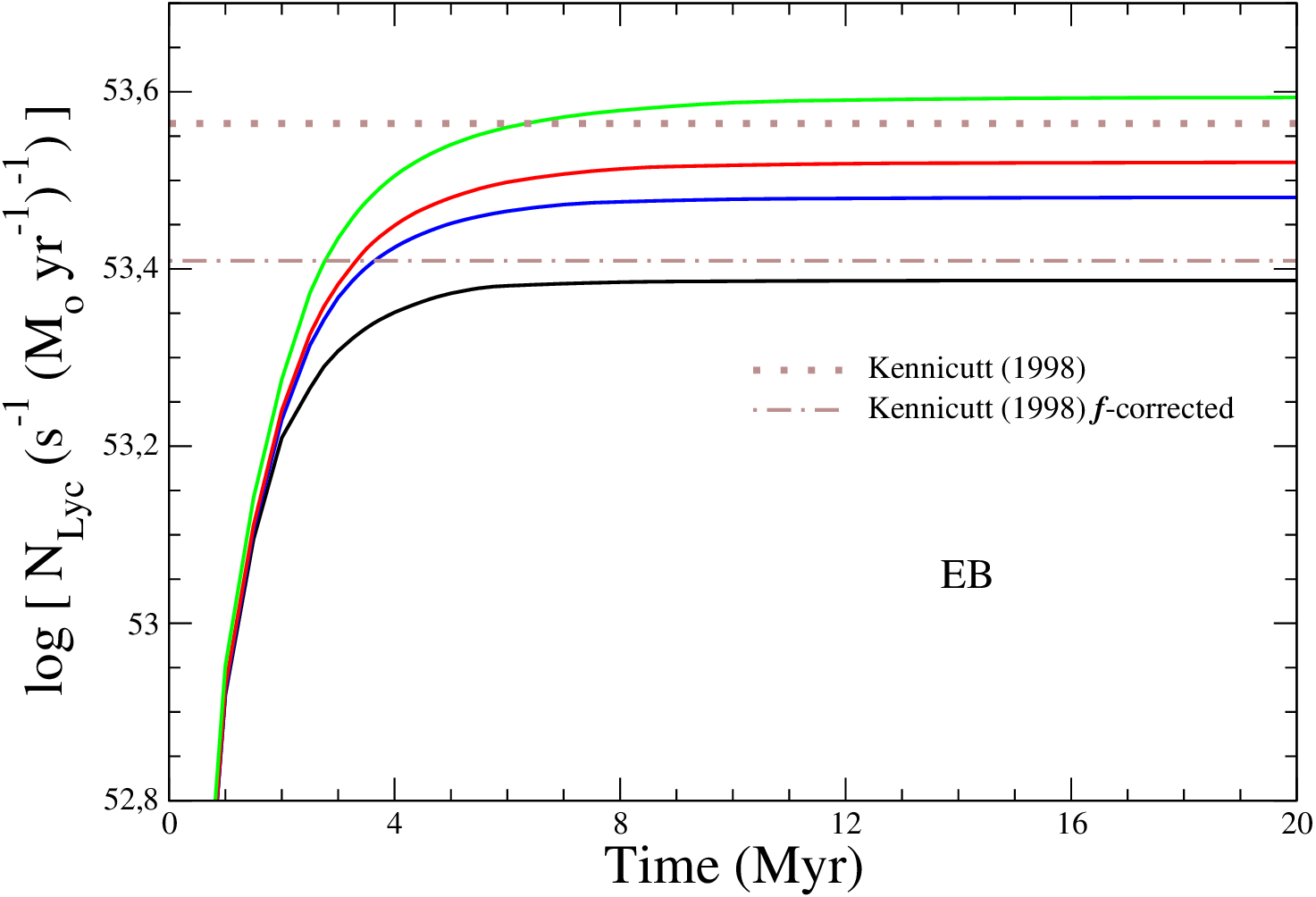}
\end{center}
\caption{Evolution of \nlyc\ predicted by CMHK02 for IB models (left, normalized to \sfs\ $= 1$ \msun) and for EB models (right, normalized to \sfr\ $= 1$ \msun\ yr$^{-1}$) as a function of metallicity (from bottom to top: $Z=0.020\mbox{-black, }0.008\mbox{-blue, }0.004\mbox{-red, }0.001\mbox{-green}$), together with the $f$ factor corrected and non-corrected \nlyc\ values from \citet{Kennicutt98}, both adapted to the IMF used in this work.} 
\label{fignlyc}
\end{figure*}

\subsection{Far infrared luminosity  (\lfir)}
\label{infrared}
The FIR emission predicted by CMHK02 models is calculated with the assumption that interstellar dust remains in thermal equilibrium. All energy absorbed, mainly stellar UV continuum, is therefore reradiated in the FIR range ($1-1000 ~\mu\rm{m}$). \citet{Calzetti2000} noticed that this parametrization is higher by up to $75$\% than the value obtained using the expression by \citet{Helou88}, which models the IR output within $42.5-122.5 ~\mu\rm{m}$ via the {\em IRAS} fluxes at $60 ~\mu\rm{m}$ and $100 ~\mu\rm{m}$, assuming a single dust component with $T=20-80$ K. The Galactic extinction law by \citet{Cardelli89} with $R_{V}=3.1$, is applied to the synthetic spectral energy distributions calculated by the models to derive \lfir. The extinction is parameterized by the value of the color excess \ebv.  Since \lfir\ saturates for  \ebv\ $>0.5$, the value computed for \ebv $ = 1.0$ should be taken as the maximum FIR emission expected in the starburst \citep{MasHesse08}. We slightly modified the extinction law, assuming that photons with a wavelength of $912$ \AA\ $< \lambda < 1250$ \AA\ suffered the same extinction $A_{1250\;\AA}$. In this way, we both a) take into account photons with $912$ \AA\ $< \lambda < 1000$ \AA, which would otherwise be ignored, and b) use a more realistic $A_{\lambda}$ value, bearing in mind that the attenuation law does not seem to increase in this range (as Eq. (5) in \citet{Cardelli89} would imply), but to flatten, as concluded by \citet{Mezger1982}. The effect of including the photons with $912$ \AA\ $< \lambda < 1000$ \AA\ in the FIR calculation is that \lfir\ is increased by $5$\%. Similarly, the computed \lfir\ changes by up to 7\% due to the modifications performed in the extinction law in the range of $912$ \AA\ $< \lambda < 1250$ \AA, but this is only detectable for low \ebv\ values. A fraction $1-f=0.3$ of ionizing photons, irrespective of their energy, is assumed to be absorbed by dust before they can ionize any atom, as derived  by  \citet{Mezger1978} and \citet{Degioia1992} and recommended by \citet{Belfort1987}. The hypothetical presence of totally obscured stars was not considered. Since no assumption is made about the temperature of the dust, the models do not yield the infrared spectrum, but only the total energy absorbed by dust, which would be reemitted in the FIR range.

The same prescriptions as those considered in CMHK02 were taken with the SB99 models to model the FIR emission of the burst. The extinction law by \citet{Cardelli89}, with the minor changes explained above, was applied to the predicted SEDs, and $30$\% of the ionizing radiation was assumed to be converted into FIR emission. The estimation of \lfir\ using the SB99 models is $\sim7$\% higher than predicted by CMHK02 for the same conditions, apparently due to the differences in the stellar atmospheres used, especially in the Lyman continuum .

\begin{figure*}
\begin{center}
\includegraphics[width=8cm,bb=17 40 714 522
dpi,clip=true]{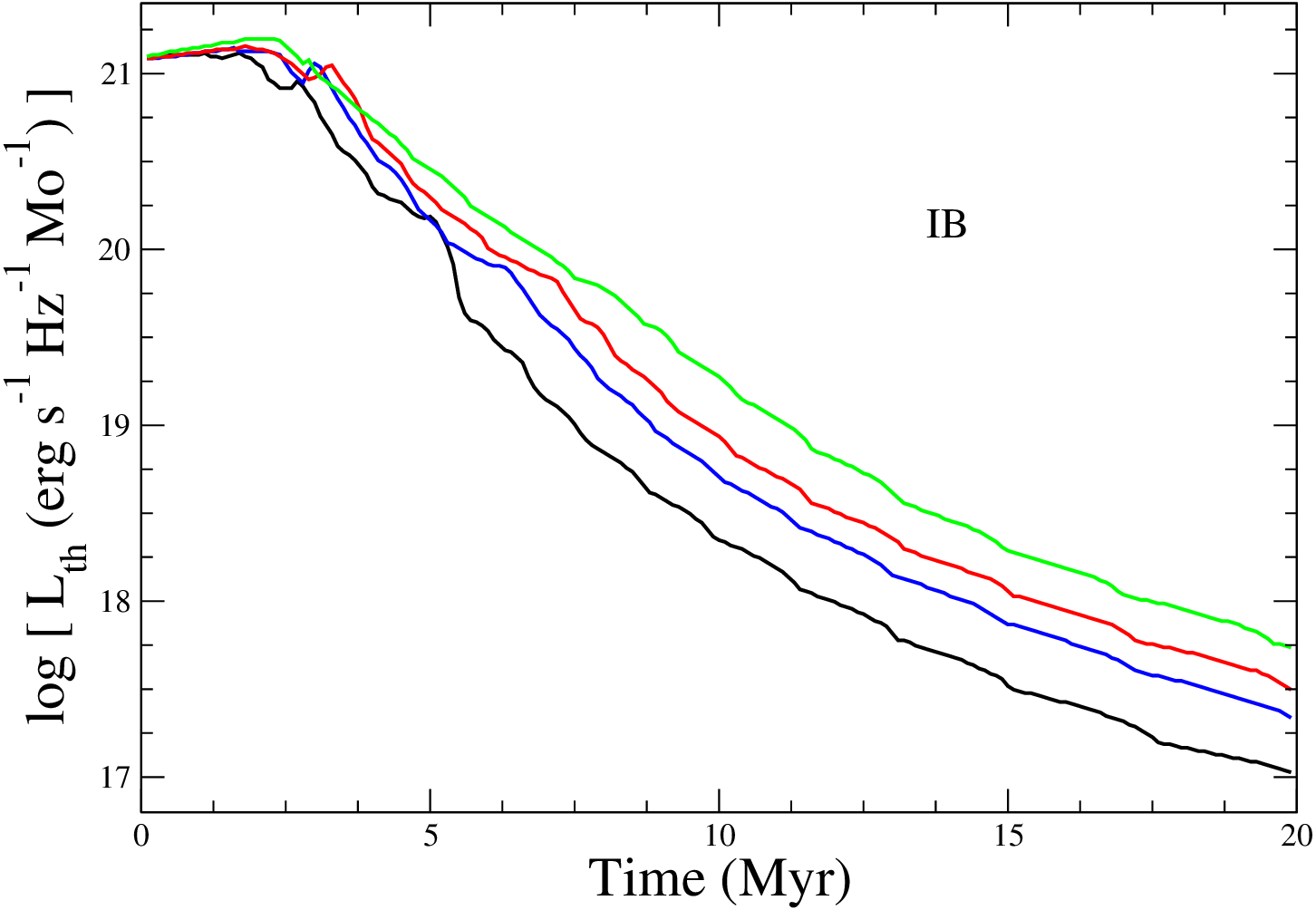}
\hspace{1.0 cm}
\includegraphics[width=8cm,bb=5 40 714 522
dpi,clip=true]{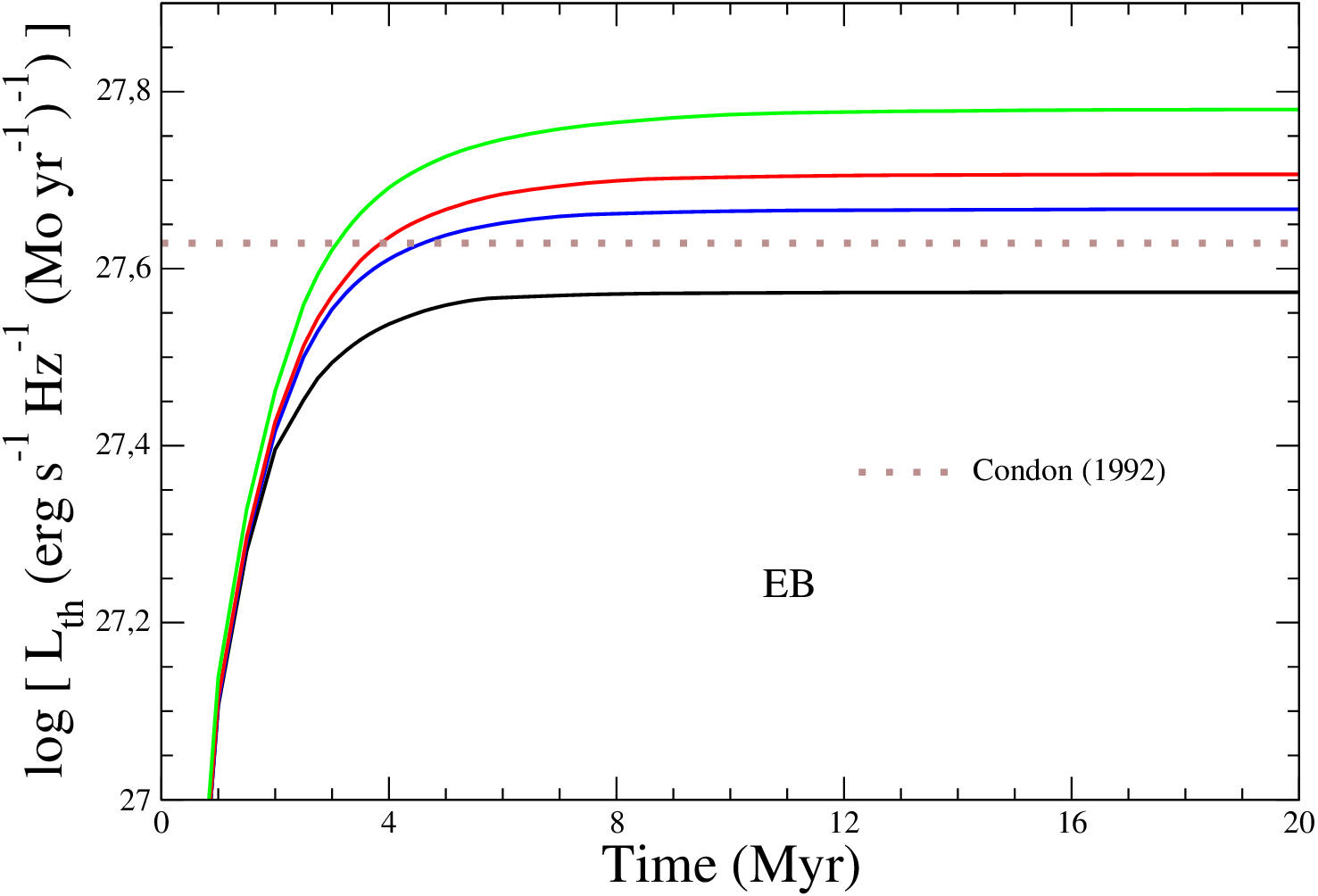}
\end{center}
\caption{Evolution of \lth\ predicted by CMHK02 for IB models (left, normalized to \sfs\ $= 1$ \msun) and for EB models (right, normalized to \sfr\ $= 1$ \msun\ yr$^{-1}$) as a function of metallicity (from bottom to top: $Z=0.020\mbox{, }0.008\mbox{, }0.004\mbox{, }0.001$). The dotted line shows the prediction by \citet{Condon92} for EB models, adapted to the IMF used in this work.} 
\label{figradioth}
\end{figure*}

\begin{figure*}
\begin{center}
\includegraphics[width=8cm,bb=5 40 714 529
dpi,clip=true]{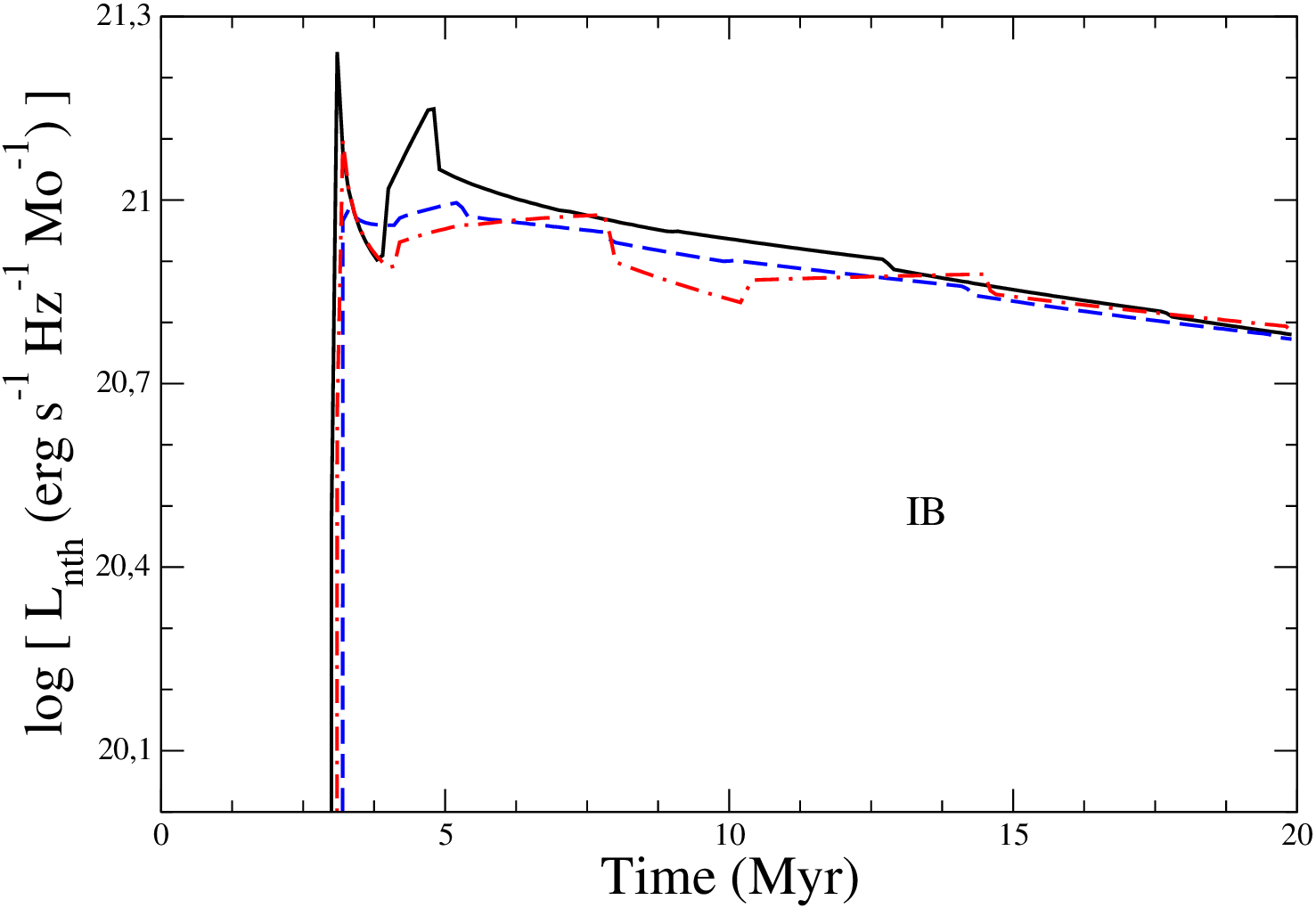}
\hspace{1.0 cm}
\includegraphics[width=8cm,bb=5 40 718 522
dpi,clip=true]{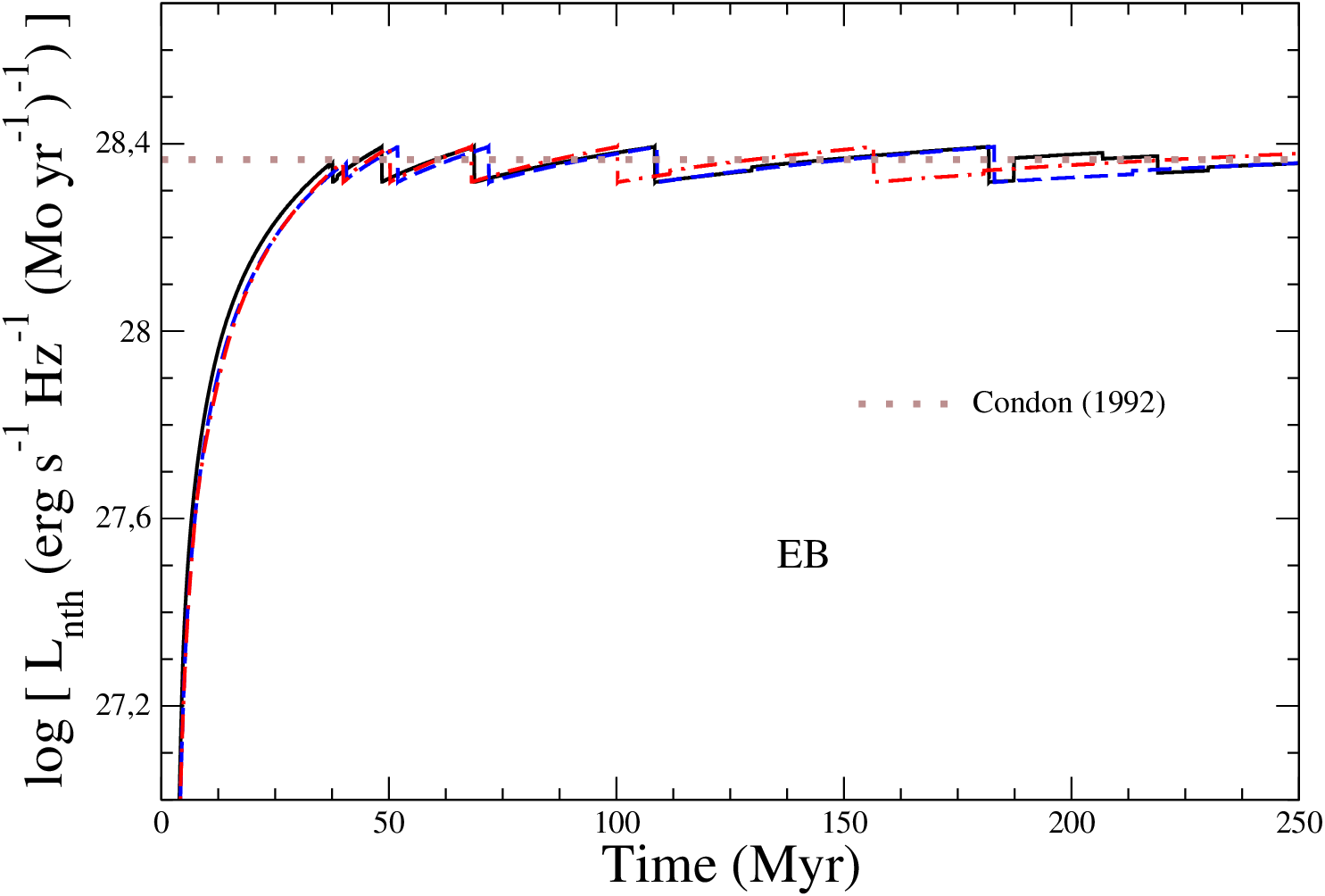}
\end{center}
\caption{Evolution of \lnth\ predicted by CMHK02 for IB models (left, normalized to \sfs\ $= 1$ \msun) and by SB99 for EB models (right, normalized to \sfr\ $= 1$ \msun\ yr$^{-1}$) for the metallicity values $Z=0.020\mbox{ (solid line), }0.008\mbox{ (dashed line), }0.001$ (point-dashed line).The dotted line shows the prediction by \citet{Condon92} for EB models, adapted to the IMF used in this work.} 
\label{figradionth}
\end{figure*}

\subsection{Ionizing power (\nlyc) and emission lines}
\label{lines}
As discussed above, we considered that a fraction $1-f=0.3$ of the ionizing photons emitted by the stars is directly absorbed by dust, so that  \nlyc\ represents only the number of photons which eventually could ionize the atoms in the surrounding gas. Since the intensities of the recombination lines are directly proportional to \nlyc\ \citep{Osterbrock89}, we wish to remark that our predictions are $30$\% lower than the luminosities computed by models which do not take this effect into account, which otherwise is realistic and should be included. Nevertheless, for the SC02 Pop. III predictions we have considered that no ionizing photons are destroyed, since at this extremely low metallicity no dust particles should as yet be present.

The intensity of the emission lines has been derived from the predicted \nlyc\ value, assuming Case B conditions ($T_{e}=10^{4}$ K,    $n_{e}=500$ cm$^{-3}$) and the following relations \citep{Storey95}:

\lha$= 1.36 \times 10^{-12} $~\nlyc\ \ergs 

\lha$= 2.86 $~\lhb

\lulyalp$ = 8.7$~\lha.

We have computed the intensities of the emission lines for various \ebv\ values, but we have not considered the effects of radiation transfer on \lyalp\ photons.  It is well known that resonant scattering by neutral hydrogen significantly affects the \lyalp\ emission line profile and can lead to the total suppression of the line \citep{Verhamme06}, depending on the column density and kinematics of the neutral gas and the amount of dust \citep{Kunth98, MasHesse03}. The \lyalp\ luminosity we have computed therefore corresponds to the upper value expected to escape from the star-forming region, corrected for the interstellar extinction in the same way as the Balmer lines, but not affected at all by neutral hydrogen scattering. In order to derive the star formation rate from the observed \lyalp\ luminosity this effect should be corrected before applying the calibration. It is important to note that the \lyalp\ photons escape fraction can cover a wide range, from 0 to 1, as shown by \citet{Atek09}, though in nearby starburst galaxies the escape fraction is close to 10\% \citep{Ostlin08}, and around 5\% in average in star-forming galaxies at $z \sim 2$ \citep{Hayes09}.

\subsection{Continuum luminosity}

We have computed the luminosity evolution of the burst's continuum emission at several wavelengths within the UV, optical and IR spectrum: $1500$ \AA, $2000$ \AA, $3500$ \AA\ (U), $4400$ \AA\ (B), $5500$ \AA\ (V) and $2.2$ $\mu$m \AA\ (K). We included both the stellar and nebular continuum components and have considered the effect of interstellar extinction as parameterized by \ebv. The most massive stars dominate the UV emission of star-forming regions and therefore drive the evolution of \lcuva\ and \lcuvb.  On the other hand, the contribution of less massive stars can become important at longer wavelengths. In EB models, medium/low-mass stars accumulate due to their longer lifetimes and become the dominant contributors to \lgb, \lgv\ and \lgk\ after $\sim 100$ Myr of evolution. 

\begin{figure*}
\begin{center}
\includegraphics[width=8cm,bb=5 40 714 528
dpi,clip=true]{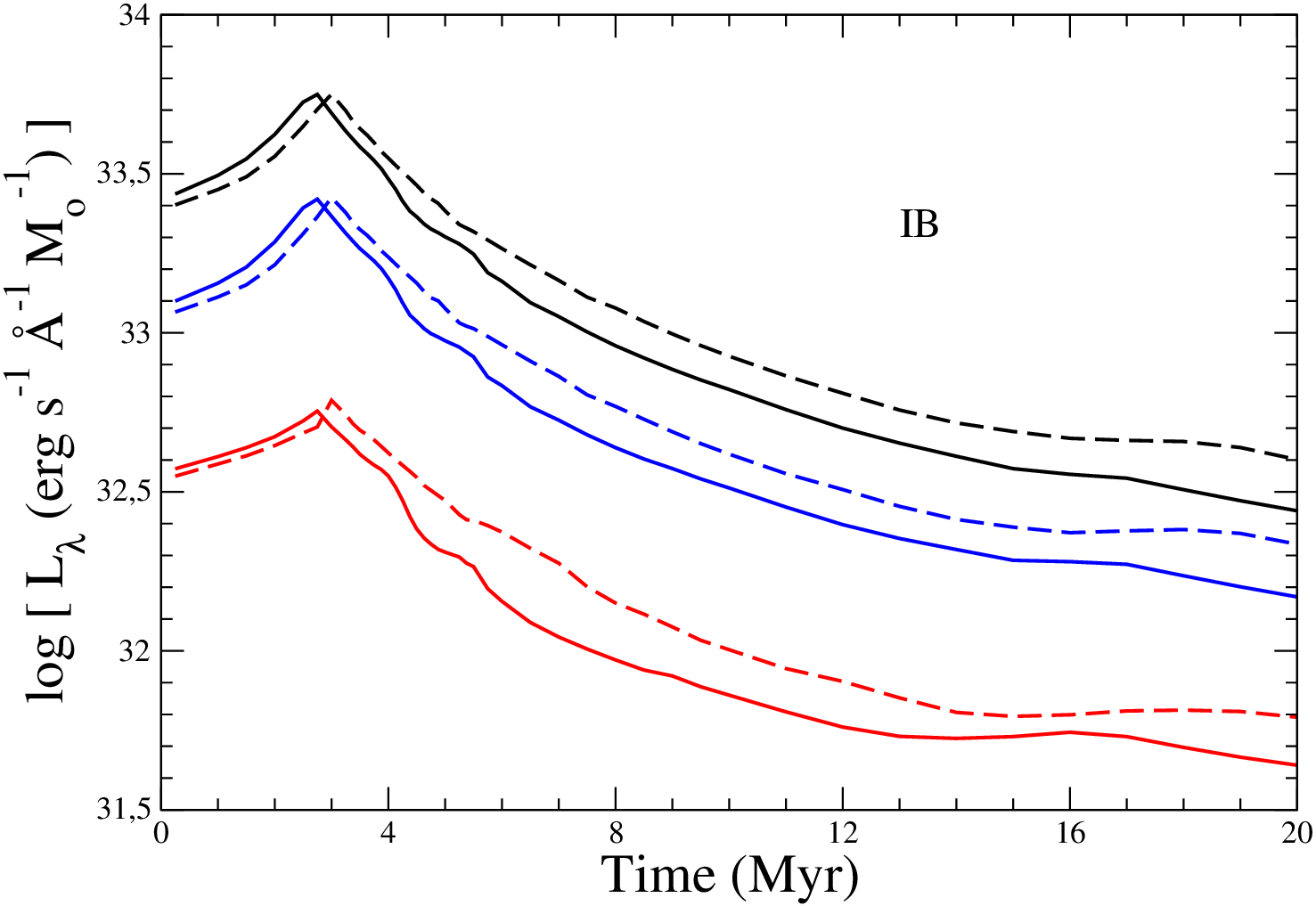}
\hspace{1.0 cm}
\includegraphics[width=8cm,bb=5 40 718 522
dpi,clip=true]{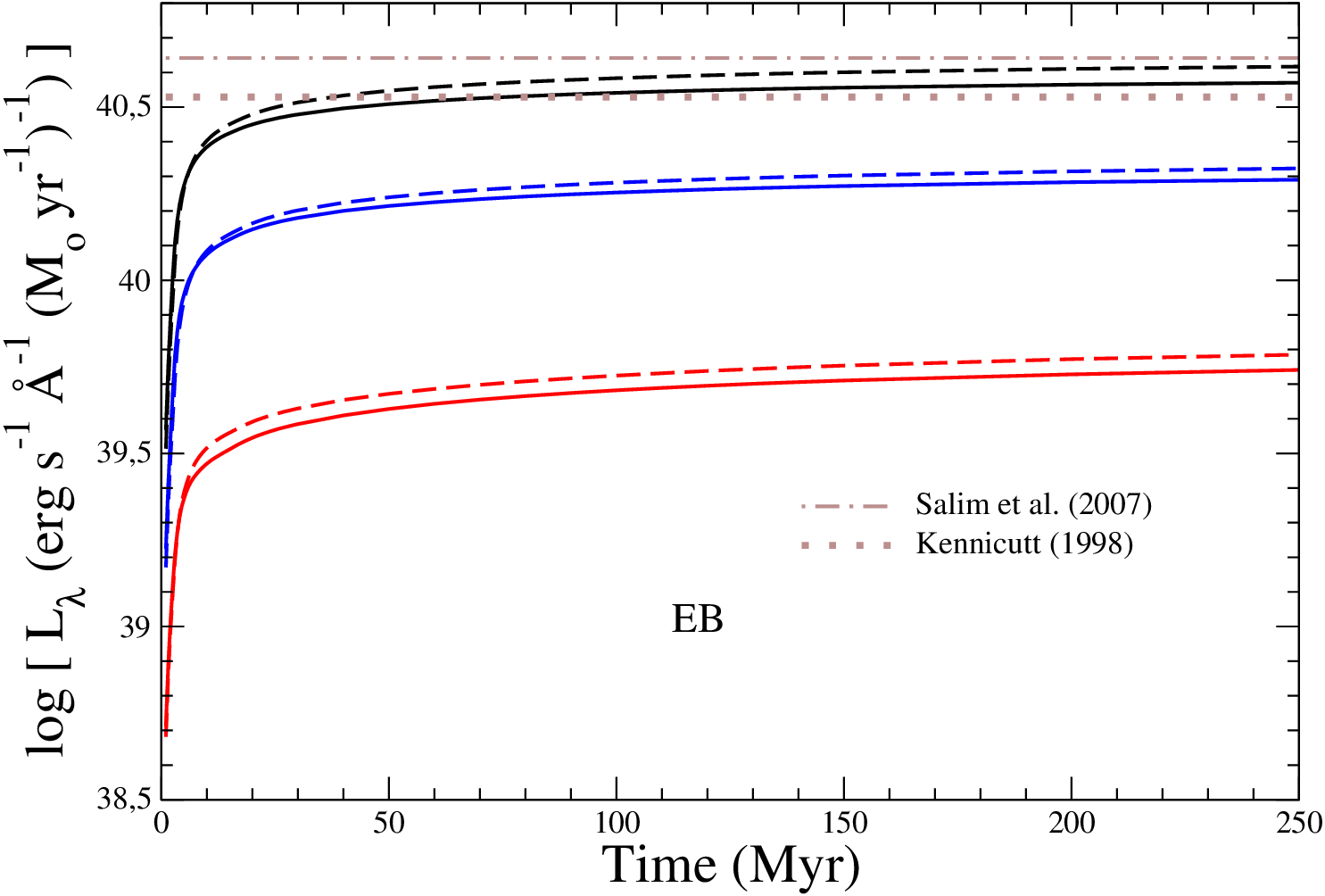}
\end{center}
\caption{Evolution (from top to bottom) of \lcuva, \lcuvb\ and \lgu\ predicted by CMHK02 for IB models (left, normalized to \sfs\ $= 1$ \msun) and by SB99 for EB models (right, normalized to \sfr\ $= 1$ \msun\ yr$^{-1}$) both for $Z=0.020$ (solid lines) and $Z=0.008$ (dashed lines). Also in EB graph, values for \lcuva\ from the calibrations by \citet{Kennicutt98} and \citet{Salim07}, both adapted to the IMF used in this work.} 
\label{figuvu}
\end{figure*}

\subsection{X-ray luminosity (\lx)}

The X-ray luminosity associated to a massive star formation episode is directly related to the amount of mechanical energy released into the interstellar medium (ISM) by stellar winds and supernova explosions. In order to calculate the mechanical energy relased by supernova explosions, both the CMHK02 and SB99 models assume that each SN injects $10^{51}$ erg into the ISM, but CMHK02 models subtract the energy emitted as X-ray radiation by SNR afterwards from the total energy of the supernova. See \citet{Cervino02} for a more detailed description. Massive early-type stars, and especially Wolf-Rayet stars, lose a significant fraction of their mass in the form of strong winds. These winds end up interacting with the interstellar gas, which becomes extremely heated. As a result, soft X-ray emission is produced. Both models use the expressions of the gas terminal velocity from \citet{Leitherer92} together with their own prediction of mass loss to calculate the mechanical energy released by the winds. 

We have computed the evolution of the mechanical energy released by the burst per unit time, \emech. This energy is converted into X-ray luminosity assuming an efficiency factor \xeff. As discussed by \citet{MasHesse08} the soft X-ray luminosity of star-forming galaxies can be reproduced by our synthesis models assuming \xeff\ in the range of $1 - 10$\%. We assumed \xeff $=0.05$ for the calibrations presented in this work. The total soft X-ray luminosity is then obtained by adding the contribution by the SNR present in the star-forming region, still active as X-ray emitters.

\subsection{Radio luminosity (\lradio)}

Both radio components, thermal and non-thermal, were calculated at $1.4$ GHz. For the thermal emission \lth we followed the prescriptions by \citet{Lequeux81}. This thermal component of the radio emission is directly proportional to the ionizing power of the starburst, with a slope \alphath\ $=-0.1$:

\noindent \lth\ $=1.57\times 10^{-26} (\nu(\rm{GHz}))^{-0.1} (T_{e}/10^{4} \rm{K})^{0.34} $\nlyc\ (\ergs\ Hz$^{-1}$).

We assumed an electronic temperature $T_{e}=10^{4}$ K and a frequency $\nu = 1.4$ GHz (i.e., 21.4~cm).

The models by \citet{MasHesse91} underestimated the non-thermal emission associated to a starburst region. \citet{MasHesse92} showed that for the ratio \lfir$/$\lradio\ to remain rather constant during the evolution of the burst, so that the observational well-known correlation between radio and FIR emission could be explained, the total non-thermal emission observed should be $\sim10$ times larger than computed. It was argued that interactions between individual supernova remnants would yield a larger radio emission than observed in isolated SNR in our Galaxy, as the ones used for the standard calibration. The same phenomenon was found by \citet{Condon90} when they applied their Eq. (7) to our Galaxy, stating that the disagreement arises because the effect of accelerated electrons after $\sim2\times10^{4}$ yr is ignored. They obtained their Eq. (8), which seems to overcome this problem. We used the expression 

\lnth\ $=1.2\times 10^{30} (\nu(\rm{GHz}))^{-0.9} \nu_{SN}$ (\ergs\ Hz$^{-1}$),

where $\nu_{SN}$ is the supernova rate. This expression is similar to Eq. (8) from \citet{Condon90}, but assumes \alphanth\ $=-0.9$ instead of \alphanth\ $=-0.8$. For consistency, the factor used was $1.2$ instead of $1.3$ to reproduce the Galactic values for \lnth\ at $408$ MHz and $\nu_{SN}$ from \citet{Berkhuijsen84} and \citet{Tammann82}, respectively. The \lnth\ value at 1.4~GHz  obtained with this expression is therefore 11\% lower than that from \citet{Condon90}. 

We have finally calibrated both \sfs\ and \sfr\ based on the total radio luminosity at $1.4$ GHz \lradio\ = \lth\ $+$ \lnth.

\section{Results and discussion} 

\label{imfcorr}

In Tables \ref{estimatorsIBall} to \ref{estimatorsEB250} we present the calibration of the different estimators computed with our models. We list the values of the factor $C_{A}$ by which the different observables must be multiplied to obtain either \sfs\ or \sfr, i.e. $SFS(L_{A}) = C_{A} L_{A}$, $SFR(L_{A}) = C_{A} L_{A}$. As discussed above, \citet{MasHesse08} showed that it is very important to take into account the evolutionary state of the burst when estimating its strength, especially in the case of IB regimes, but also in some cases for EB episodes. For IB models we have calibrated the estimators for ages of $4$, $5$ and $6$ Myr, which is the typical range of ages measured for star-forming galaxies \citep{MasHesse99b, Pellerin07}. The ionizing power decreases very rapidly, so that the emission lines are barely detectable after 7~Myr: \citet{MasHesse99b} predict \ewhb\ values below 2~\AA\ after this age for solar metallicity starbursts.

On the other hand, in an EB regime stars are replaced by new massive stars as they die, eventually reaching a steady state after a few tens of Myr. For this reason, we studied a wider range of ages: a non-evolved continuous episode with $10$ Myr, a young burst after $30$ Myr of evolution, and a star formation process already in the steady state, at $250$ Myr. Not all parameters remain constant when the steady state is reached, since medium/low-mass stars accumulate and contribute more and more to the optical -- IR stellar continuum. As a result the calibration of some tracers is a function of the evolutionary state as well, though the effect is much weaker than for IB regimes. For \nlyc\ and hence for \lha\ and \lulyalp, the asymptotic state is reached very soon, so that the same $C_{A}$ value is displayed for the three points at $10$ Myr, $30$ Myr and $250$ Myr. 

For the parameters which are affected by interstellar extinction we computed the calibration of the different estimators for different \ebv\ values within the range \ebv$= 0.0 - 1.0$ under the assumption of the Galactic extinction law by \citet{Cardelli89}.

As mentioned above, our calibrations have been computed for a Salpeter initial mass function with limits at $2$ and $120$ \msun. To ease the comparison with other calibrations we list the correction factors in Tables \ref{corfactorsIB} and \ref{corfactorsEB} by which values from Tables \ref{estimatorsIBall}-\ref{estimatorsIB06}, and \ref{estimatorsEBall}-\ref{estimatorsEB250}, respectively, must be multiplied to convert our calibrations to other usual IMF mass ranges of $M=0.1-100$ \msun\ and $M=1-100$ \msun. Similarly, in Tables \ref{corfactorsIBalpha} and \ref{corfactorsEBalpha} we list the conversion factors which would be required for two limiting IMF slopes, with $\alpha = 1$ and $3$. These correction factors are the values $L(2-120,2.35)/L(M_{\rm{low}}-M_{\rm{high}},\alpha)$, where $M_{\rm{low}}$ and $M_{\rm{high}}$ are the new mass limits, $\alpha$ is the new IMF slope and $L$ the magnitude. For instance, the \sfrrad\ calibration for a burst that has been creating stars with a constant rate for $30$ Myr, and assuming an IMF with mass range $M=0.1-100$ \msun\ and $\alpha=2.35$, is \sfrrad\ $= 3.51\;4.3\times 10^{-29}$\lradio\ $= 1.5\times 10^{-28}$\lradio\ (luminosities are measured in \ergs\ Hz$^{-1}$ and \sfr\ in \msfr).  We calculated the values of this correction factor using the SB99 models, since these models allow the IMF to be defined by the user.

This correction between different IMFs depends on the age of the burst, on the parameter studied and on the star formation history assumed in the model and has two different origins: a) when considering different mass limits, both \sfr\ and \sfs\ measure the mass converted into stars in a different mass range, and b) if $M_{\rm{high}}$ or $\alpha$ changes, so does the amount and eventual type of massive stars that contribute to the emission, which can be significantly affected. Emission by low-mass stars does not contribute dramatically to any of the computed parameters at the ages considered, and therefore any change in $M_{\rm{low}}$ only exerts a variation of type a) in the calibration, as long as it does not increment unphysically. Also, since stars with masses above $100$ \msun\ have already died in IB models at the ages considered in the calibrations, only correction of type a) is needed (if $\alpha$ is not modified), which can be performed analytically by the calculation of the ratio of the integrated IMFs \citep{Schaerer03,Wilkins08}, and which does not depend on the evolutionary state. On the contrary, in EB models there are always stars more massive than $100$ \msun, hence a proper correction for both a) and b) using SB99 as explained is needed. As can be checked in Tables \ref{corfactorsIB} and \ref{corfactorsEB}, and also by a comparison of these values with those from \citet{Schaerer03}, values of correction factors obtained analytically agree with those obtained for EB models using SB99 typically within $5$\%, with the difference being higher only for \nlyc\ ($\sim10-15$\%) due to the fact that it is solely dominated by the most massive stars of the burst. Since SB99 models do not yield \lsx, it was not possible to follow the same procedure in this case as for the other magnitudes, and therefore only the analytical value for \lsx\ is given in Table \ref{corfactorsEB}.

The effect of the age of the burst in the correction of \sfr\ estimators when different mass limits are assumed is very low ($\sim8$\% at most) for the ages considered, hence we decided to leave only the asymptotical value at $250$ Myr in Table \ref{corfactorsEB}. Similarly, the effect of \ebv\ value in the \sfrfir\ correction factor is typically below $5$\%, therefore we opted for showing only the correction obtained for \ebv\ $=0.3$, which lies somewhat in the middle of the range studied, where \lfir\ has not attained the saturation described in Sect. \ref{infrared}. Since \lha\ and \lulyalp\ are proportional to \nlyc, only the correction for the latter was included. We have applied these corrections when we compared our calibrations with the results of \citet{Kennicutt98} and \citet{Salim07}, which were computed for a Salpeter IMF, but with mass limits of $0.1-100$ \msun. \citet{Condon92} assumed mass limits of $5-100$ \msun, but an IMF slope $\alpha=2.50$, therefore $L(2-120,2.35)/L(5-100,2.50)$ was calculated explicitly in order to compare his radio calibrations with our predictions.

Our calibration of star formation estimators at different energy ranges (from X-rays to radio) can also be used inversely (taking care of the different time constants and properties of the different tracers): for example, once the \sfr\ or \sfs\ is estimated from the UV or optical continuum luminosity, the expected emission at other ranges or emission lines can be directly computed, corrected for the effects of interstellar extinction in a consistent way. Nevertheless, the inverse use of the \sfr, \sfs\ calibrators to derive the expected luminosity at other energy ranges has to be taken with caution. They are valid only for galaxies dominated by an ongoing massive star formation process, and so they would predict only the luminosity associated to the star formation episode. This could be just a fraction of the total luminosity of the galaxy at other wavelength ranges dominated by an old underlying stellar population, like the near infrared.
We have built a Web tool to compute the \sfr\ or \sfs\ for any given estimator and at different conditions, which is publicly accessible at {\tt{\small{http://www.laeff.cab.inta-csic.es/research/sfr/}}}

In the sections below we discuss the properties and applicability of each of the \sfr\ and \sfs\ estimators, and we compare our results with previous calibrations.

\subsection{Far infrared luminosity  (\lfir)}

Evolution of \lfir\ is shown in Fig.~\ref{figlfir} for both the CMHK02 IB and SB99 EB models and different metallicities $Z=0.020\mbox{, }0.008\mbox{, }0.001$, computed for \ebv$= 1.0$. Since \lfir\  in star-forming regions is directly linked to the UV-optical  emission produced by the most massive stars of the bursts, it shows a peak in IB models at $\sim3$ Myr, followed by a steep decline afterwards. On the other hand, when a constant \sfr\ is assumed, \lfir\ starts to stabilize after the first 20 Myr of evolution, when an equilibrium between formation and destruction of massive stars is reached. Nevertheless, since medium/low-mass stars also contribute to \lfir, this parameter increases slowly but steadily as long as the star formation episode is active. As discussed by \citet{MasHesse08}, \lfir\ saturates for \ebv\ $>0.5$ since the attenuation is already so high that most of the continuum emission is absorbed.  We have calibrated \lfir\ as \sfr\ and \sfs\ estimator in Tables \ref{estimatorsIBall} to \ref{estimatorsEB250} for various \ebv\ values. If \ebv$= 1.0$ is assumed, \lfir\ becomes indeed a good estimate of the bolometric luminosity of the star formation process. 

The evolution of \lfir\ with time is sensitive to the metallicity, since low-metallicity stars evolve more slowly.  As a result,  \lfir\ is  $15-30$\% larger for  $Z=0.008$ than for solar metallicity   models at ages $4-6$ Myr.  In EB models, the effect is to accumulate more massive stars at a given time, so that \lfir\ becomes $\sim8-11$\% larger for $Z=0.008$ once the burst is older than $30$ Myr. On the other hand, during the initial phases of evolution, \lfir\ is essentially independent of the metallicity. 

Together with our predictions we show in Fig.~\ref{figlfir} the \lfir\ value derived from the expression by \citet{Kennicutt98}, which was IMF-corrected as explained above to be consistent with the IMF we have used.  The adapted  \citet{Kennicutt98} value becomes $\log$ \lfir\ $=43.90$, which agrees quite well with our  predictions for solar metallicity at ages above $\sim100$ Myr. On the other hand, the discrepancy becomes larger for less evolved bursts, differing by $\sim21$\%  at $30$ Myr, and by much more for the younger the episode.  Although \citet{Kennicutt98} affirms that his calibration applies only to bursts younger than $100$ Myr, it can be seen in Fig.~\ref{figlfir} that it could be used accurately for solar metallicity extended bursts at $50-250$ Myr.

\subsection{Ionizing power (\nlyc) and emission lines}

We show in Fig.~\ref{fignlyc} the evolution of \nlyc\ for IB and EB models at different metallicities, as predicted by CMHK02 models. The emission of ionizing photons is driven exclusively by the most massive stars in the burst. Since these stars have short lifetimes, the ionizing power of instantaneous bursts decreases drastically with time after 3~Myr, when the most massive stars begin to explode as supernovae. For the same reason, their number reaches an equilibrium value quite rapidly in models with a constant star formation rate, so that \nlyc\ reaches an asymptotic value in just $\sim8$ Myr in EB models. This steady state of \nlyc\ lasts hundreds of Myr in EB models, but  we show in the figure just the first $20$ Myr of evolution of the burst to remark its rapid stabilization. As we discussed, we considered that a fraction $1-f=0.3$ of the ionizing photons are absorbed by dust, and here we show their number which can eventually ionize the atoms in the surrounding gas.

When considering $Z < Z_{\sun}$, we find that the slower evolution of high-mass stars  translates into a delay in the emission of ionizing photons in IB models, as can be observed in Fig.~\ref{fignlyc}.  The effect is to increase \nlyc\ significantly, by $0.4$ dex at $4$ Myr and $0.7$ dex at $6$ Myr for $Z=0.001$.  In  EB models the accumulation of massive stars originates  \nlyc\ values $0.2$ dex higher for $Z=0.001$.

We include in Fig.~\ref{fignlyc}  the  \nlyc\ value by \citet{Kennicutt98}, adapted to our IMF limits. This \nlyc\ value is larger than our prediction for solar metallicity models by  $50$\%, which is apparently due to the fact that \citet{Kennicutt98} did not apply any $f$ correction.  If we add $1-f=0.3$ correction to the \citet{Kennicutt98} \nlyc\ value,  both predictions are within $6$\% in the asymptotic phase, as can be observed in Fig.~\ref{fignlyc}. We insist that, as discussed by \citet{MasHesse91}, it should be more realistic to assume that a fraction around 30\% of ionizing photons are absorbed by dust, even in relatively dust-free environments, and that they therefore do not contribute to the ionization process. 

In Tables \ref{estimatorsIBall} to \ref{estimatorsEB250}  we list the predicted \nlyc\ values for IB and EB models at different evolutionary states and the predicted \hal\ and \lyalp\ calibrations for different values of interstellar extinction, computed as discussed in Sect. \ref{lines}. In Tables~\ref{estimatorsIBpop3} and \ref{estimatorsEBpop3} we list the corresponding \nlyc\ predictions for starbursts dominated by Pop. III stars. As discussed above, $1-f=0.0$ has been assumed in this case.

We wish to remark that there may be significant differences between the extinction affecting the nebular lines and the stellar continuum. \citet{Calzetti2000} estimated that on average the color excess (\ebv) derived from the stellar continuum was a factor of 0.4 lower than  the value derived from the emission lines. The analysis of NGC 4214, a well-resolved starburst galaxy, by \citet{Maiz1998} showed that the dust appeared concentrated at the boundaries of the ionized region in its central star-forming region and affected mainly the nebular emission lines, while the stellar continuum itself was located in a region relatively free of dust and gas. But on the other hand, no spatial decoupling between massive stars, ionized gas and dust was found on other star-forming regions of the same galaxy, which indicates that the specific geometry of each star-forming knot drives the decoupling of continuum and emission line extinctions. \citet{Erb2006} concluded from the analysis of a large sample  (114 objects) of star-forming galaxies at $z\sim2$  that the average \sfr\ values derived from both \lha\ and \luv, applying the same \ebv\  for the extinction correction, were completely consistent (indeed, almost identical). Nevertheless, the scatter was significant (0.3 dex), which again indicates that the geometry of individual objects plays an important role. This effect has to be taken into account if the \sfr\ is derived simultaneously from tracers based on both the emission lines intensity and continuum level. Finally, we want to stress that the strength of the extinction assumed in our computations results from the combination of the listed \ebv\ values and the \citet{Cardelli89} law. To get the same extinction with different laws, correspondingly different \ebv\ values have to be used (for example, the same extinction is obtained at \hal\ with \ebv$= 0.3$ and the \citet{Cardelli89} law as with \ebv$ = 0.23$ and the \citet{Calzetti2000} parameterization).

\begin{figure*}
\begin{center}
\includegraphics[width=8cm,bb=5 40 714 522
dpi,clip=true]{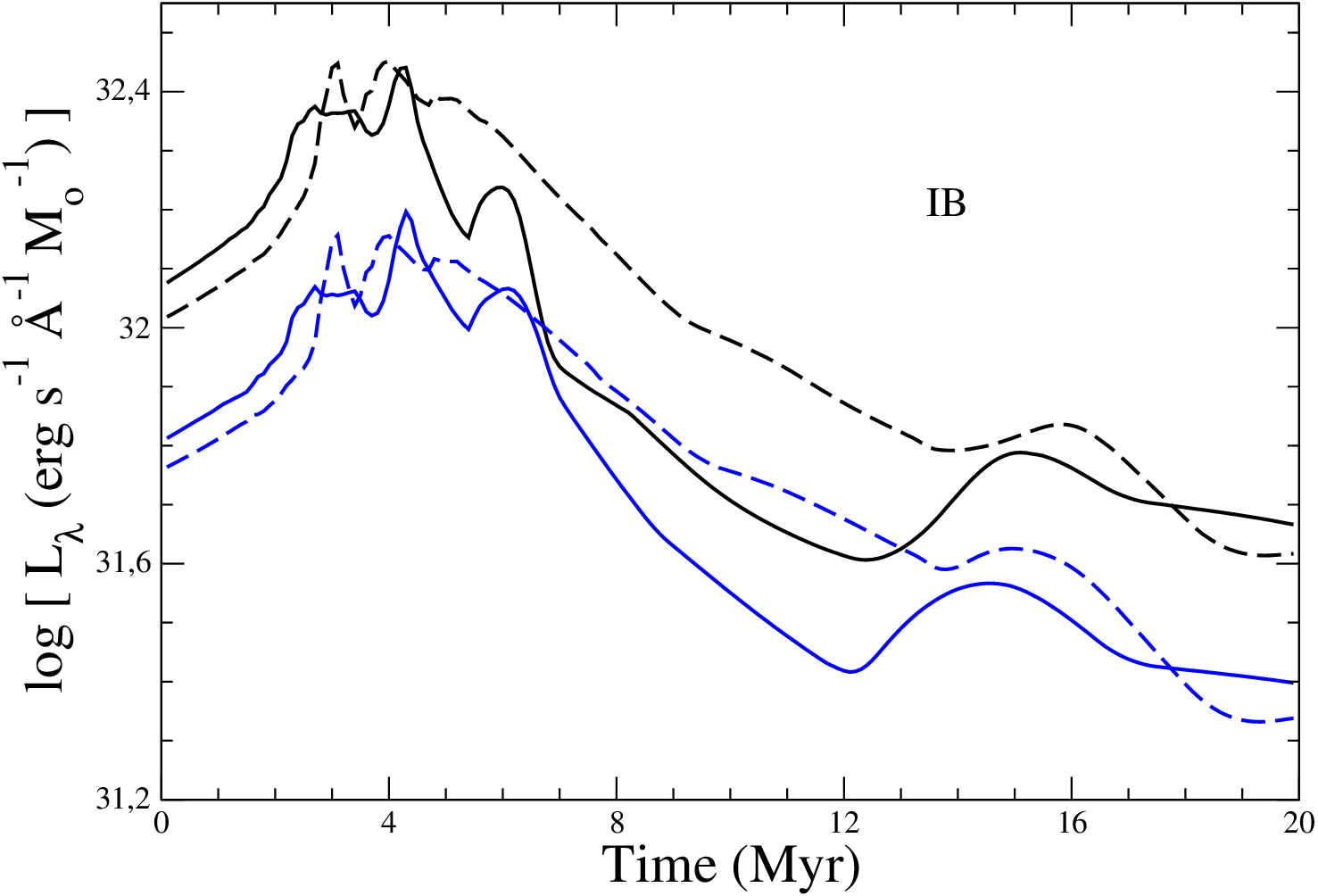}
\hspace{1.0 cm}
\includegraphics[width=8cm,bb=5 40 718 528
dpi,clip=true]{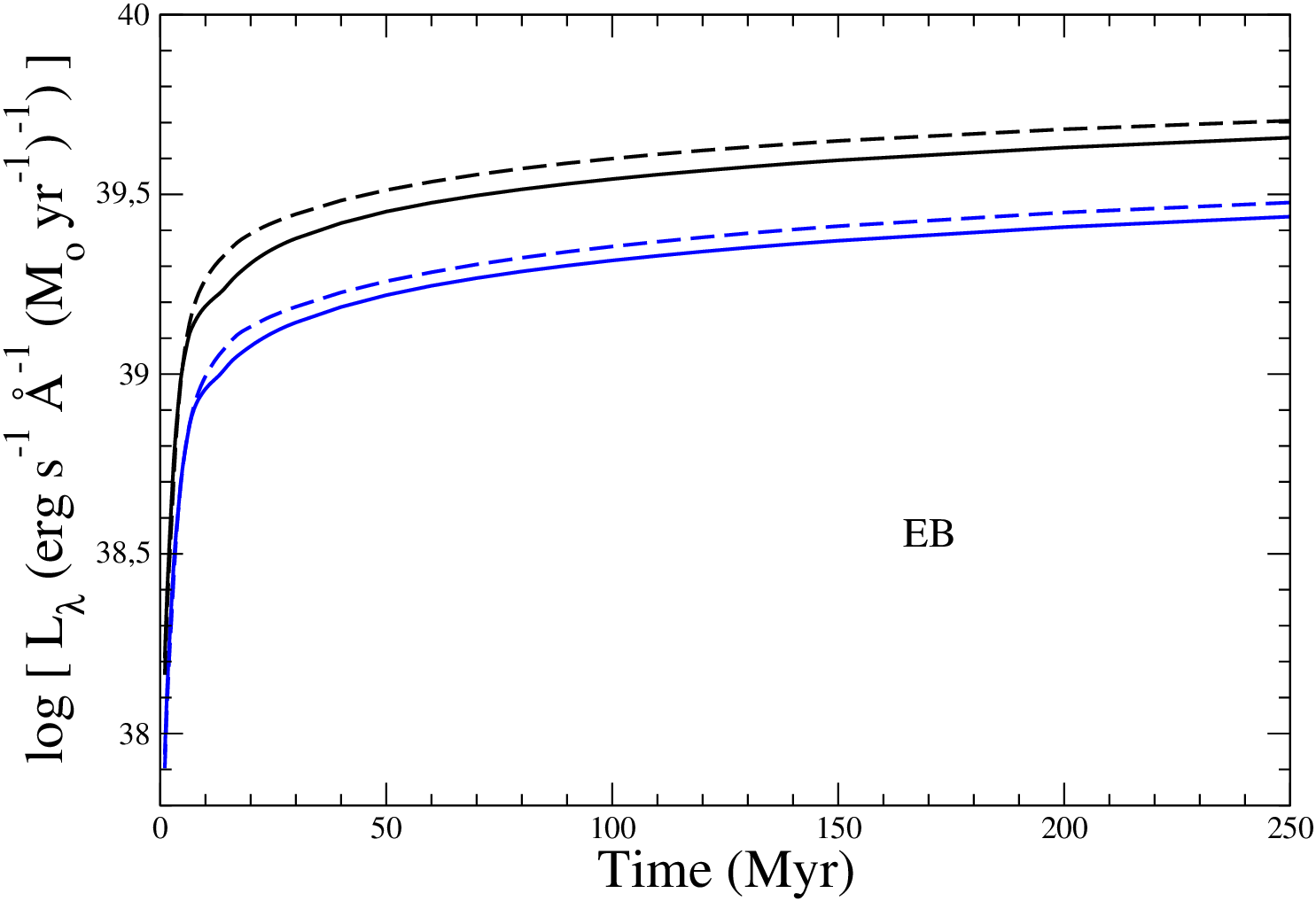}
\end{center}
\caption{Evolution (from top to bottom) of \lgb\ and \lgv\ predicted by CMHK02 for IB models (left, normalized to \sfs\ $= 1$ \msun) and SB99 for EB models (right, normalized to \sfr\ $= 1$ \msun\ yr$^{-1}$) both for $Z=0.020$ (solid lines) and $Z=0.008$ (dashed lines).} 
\label{figbv}
\end{figure*}

\begin{figure*}
\begin{center}
\includegraphics[width=8cm,bb=5 40 714 522
dpi,clip=true]{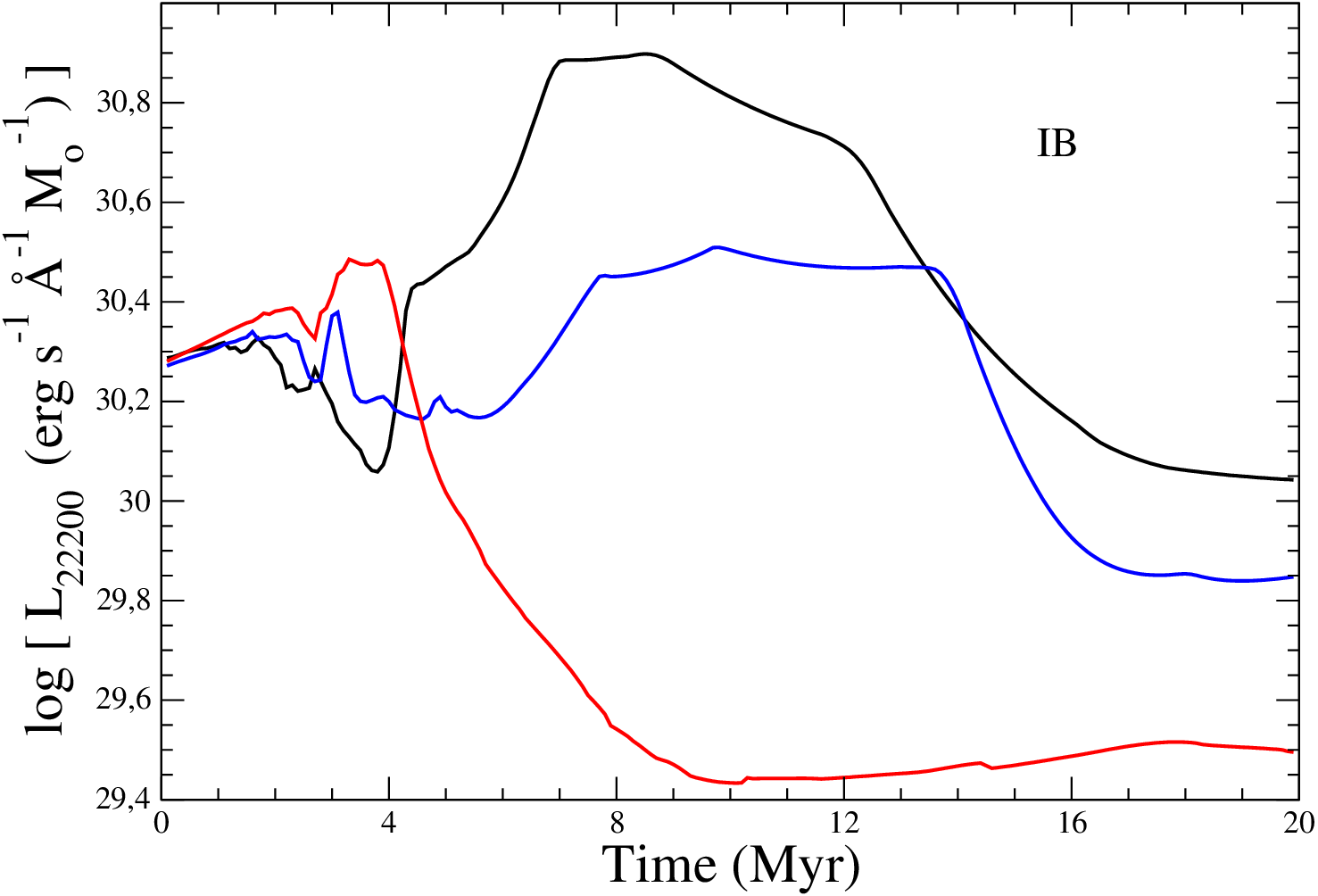}
\hspace{1.0 cm}
\includegraphics[width=8cm,bb=5 40 718 522
dpi,clip=true]{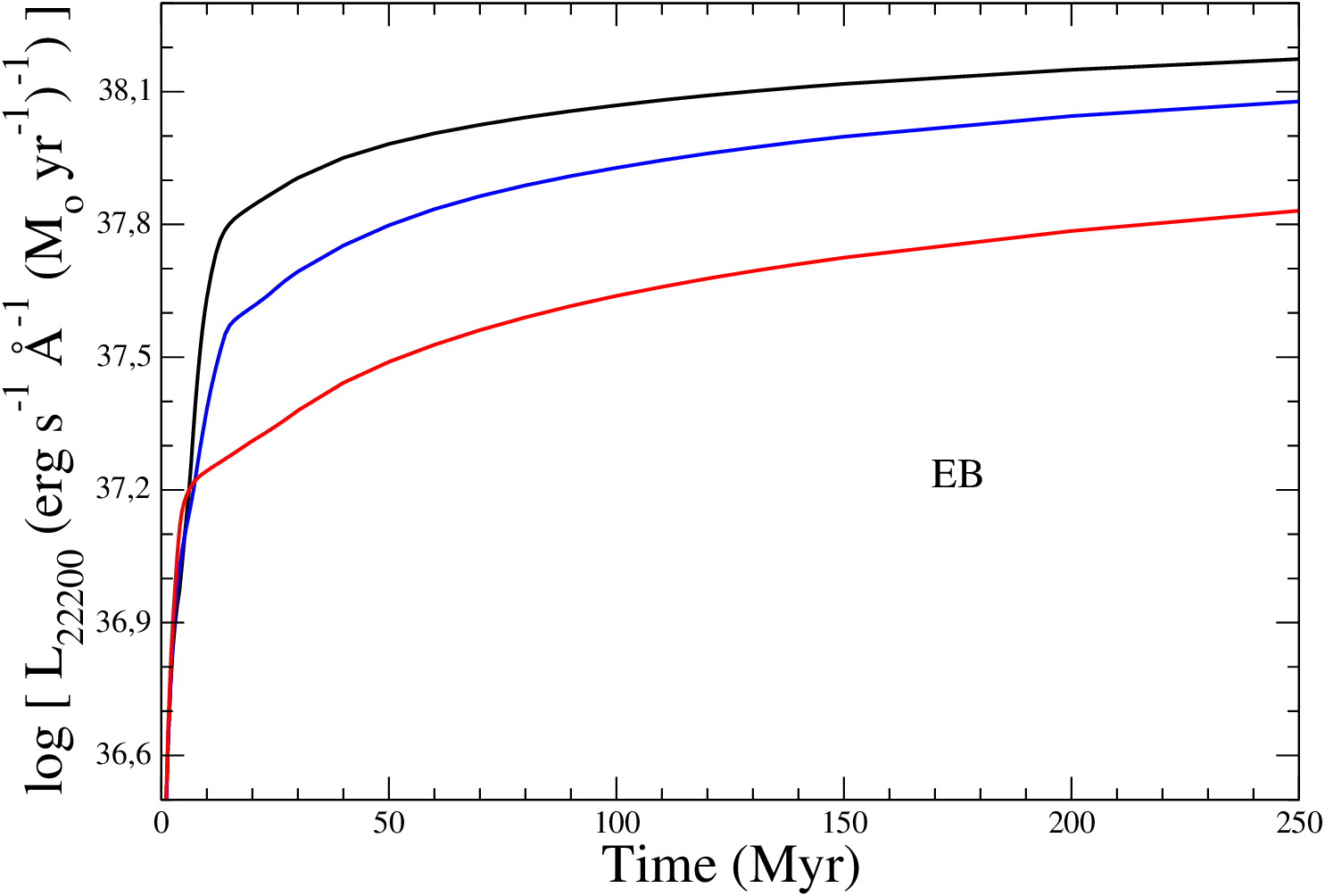}
\end{center}
\caption{Evolution of \lgk\ predicted by CMHK02 for IB models (left, normalized to \sfs\ $= 1$ \msun) and SB99 for EB models (right, normalized to \sfr\ $= 1$ \msun\ yr$^{-1}$). From top to bottom: $Z=0.020\mbox{-black, }0.008\mbox{-blue, }0.001\mbox{-red}$.} 
\label{figk}
\end{figure*}

\begin{figure}
\centering
\includegraphics[width=8cm,bb=5 40 714 522
dpi,clip=true]{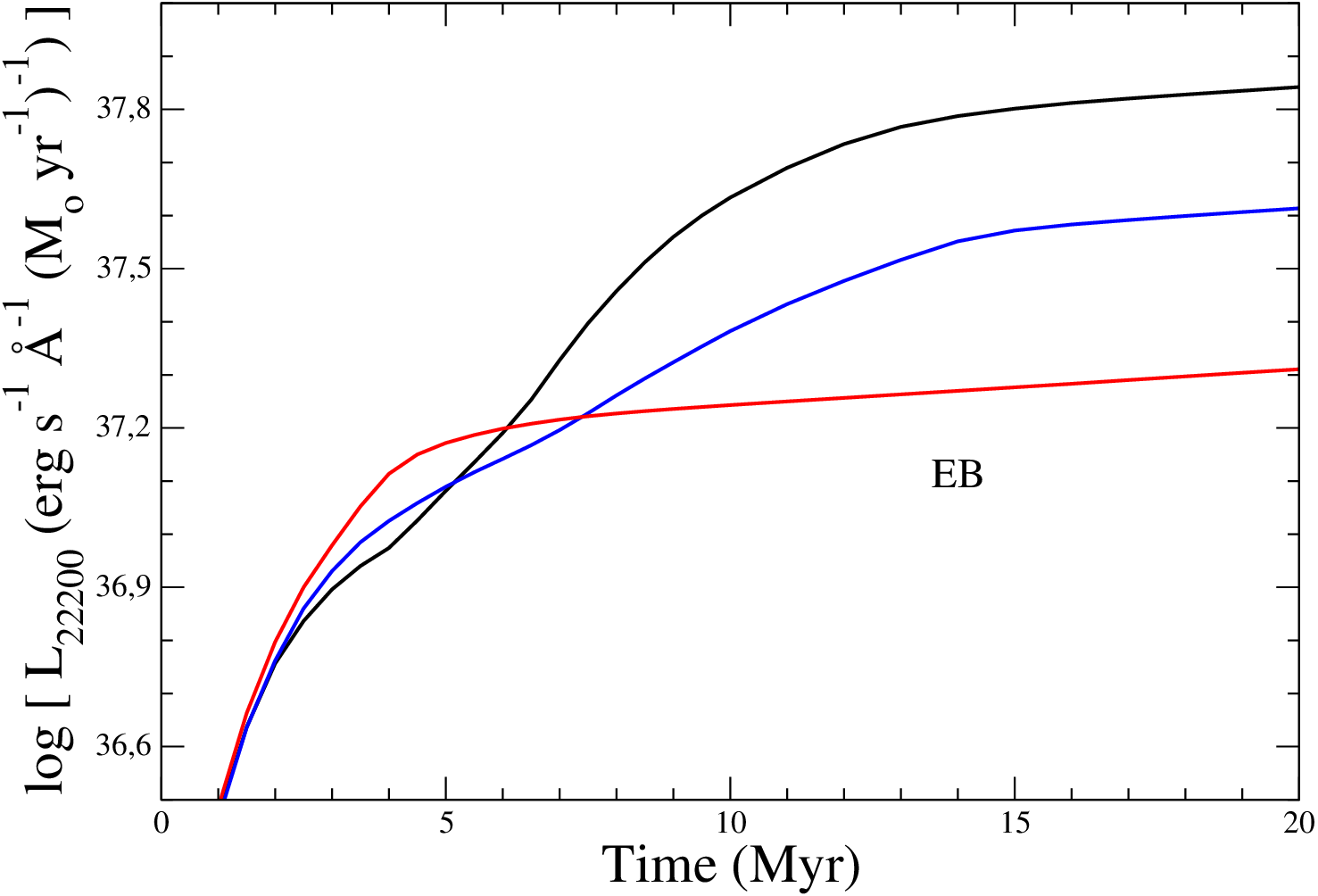}
\caption{Evolution of \lgk\ predicted by SB99 for EB models as in Fig.~\ref{figk}.} 
\label{figkEB}
\end{figure}

\subsection{Radio emission}

The evolution of \lth\ and \lnth\ is presented in Figs.~\ref{figradioth} and \ref{figradionth}, respectively. Since the thermal radio component is assumed to be proportional to \nlyc, its evolution directly follows that of \nlyc, as discussed above. But we still show the plots in Fig.~\ref{figradioth} to allow for a direct comparison between the intensities of the thermal and non-thermal components at different evolutionary states of the burst. We have also plotted on Figs.~\ref{figradioth} and \ref{figradionth} the values predicted by \citet{Condon92} for EB models after they were adapted to our IMF. 

When constant star formation is assumed, \lnth\ is observed to increase two orders of magnitude within $\sim35$ Myr after the SN start to explode, reaching then a steady state with almost no sensitivity to the metal content of the star-forming region. The predictions for IB models, on the other hand, are more sensitive to metallicity, especially at ages between 3 and 12 Myr, when the most massive stars explode (their lifetime and evolution depend on metallicity). 

In Tables \ref{estimatorsIBall} to \ref{estimatorsEB250}  we list the calibration of \lradio, which includes both the thermal and non-thermal components at any time. It can be seen in the figures that the total radio emission at 1.4 GHz is dominated by the non-thermal component as soon as supernova explosions take place in the star-forming region. Only during the first 2-3 Myr in IB bursts can the radio luminosity be dominated by thermal emission. 

Although the calibration derived by \citet{Condon92} is consistent with our predictions, Fig.~\ref{figradioth} shows that the \lth\ value is $12$\% higher than predicted by our models. However, when using the \lha\ value from \citet{Kennicutt83} as \citet{Condon92} did, with the prescriptions we have explained in Sect. \ref{lines}, it can be checked that this difference falls to $5$\%. But we wish to note that \citet{Condon92} ignores the dust absorption of ionizing photons, but assumes a $\sim30$\% lower ionizing power of the burst than predicted by the SB99 models. Both effects seem to cancel each other, which renders both \lth\ predictions consistent.

\begin{figure*}
\begin{center}
\includegraphics[width=8cm,bb=17 40 714 522
dpi,clip=true]{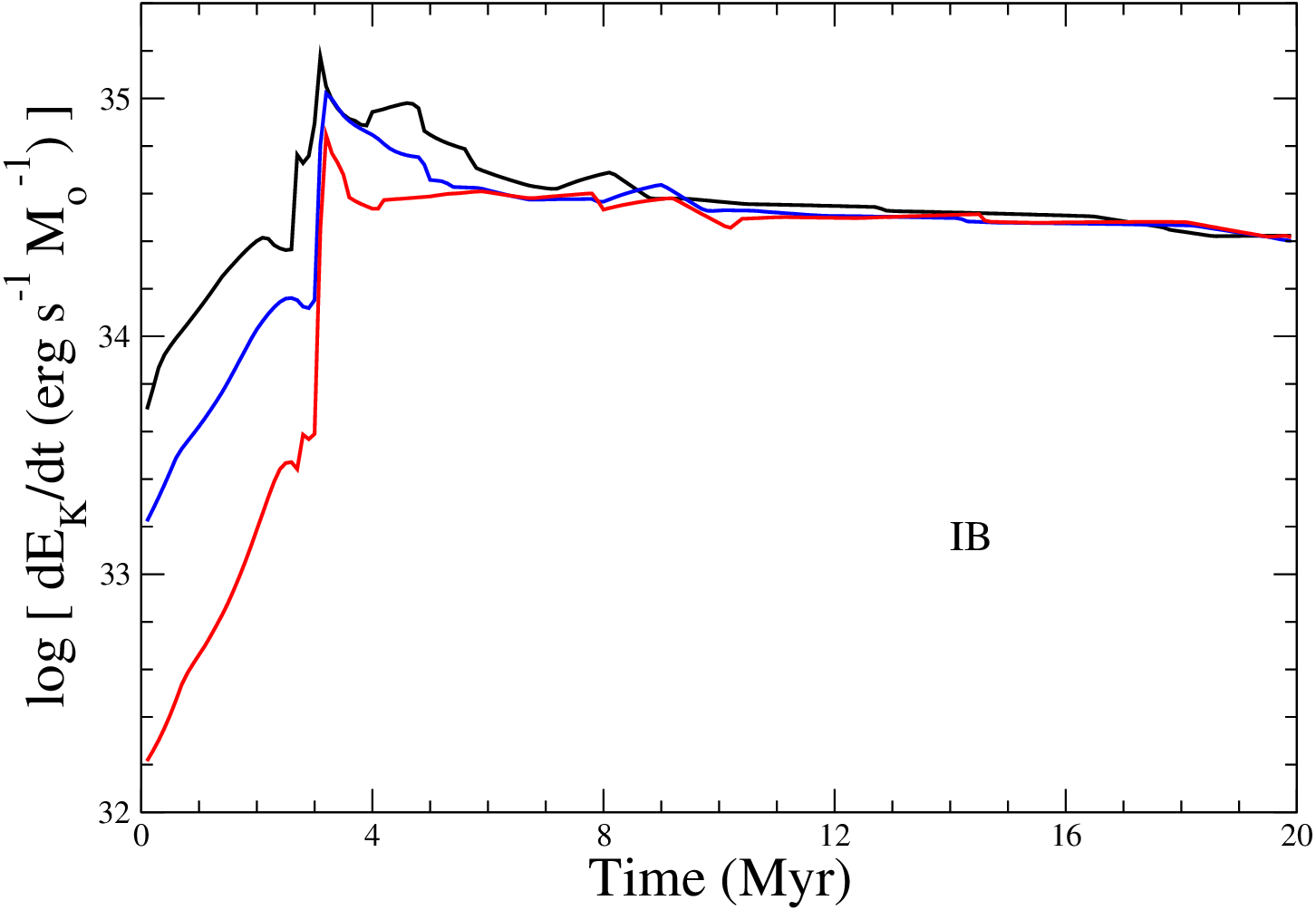}
\hspace{1.0 cm}
\includegraphics[width=8cm,bb=17 40 718 522
dpi,clip=true]{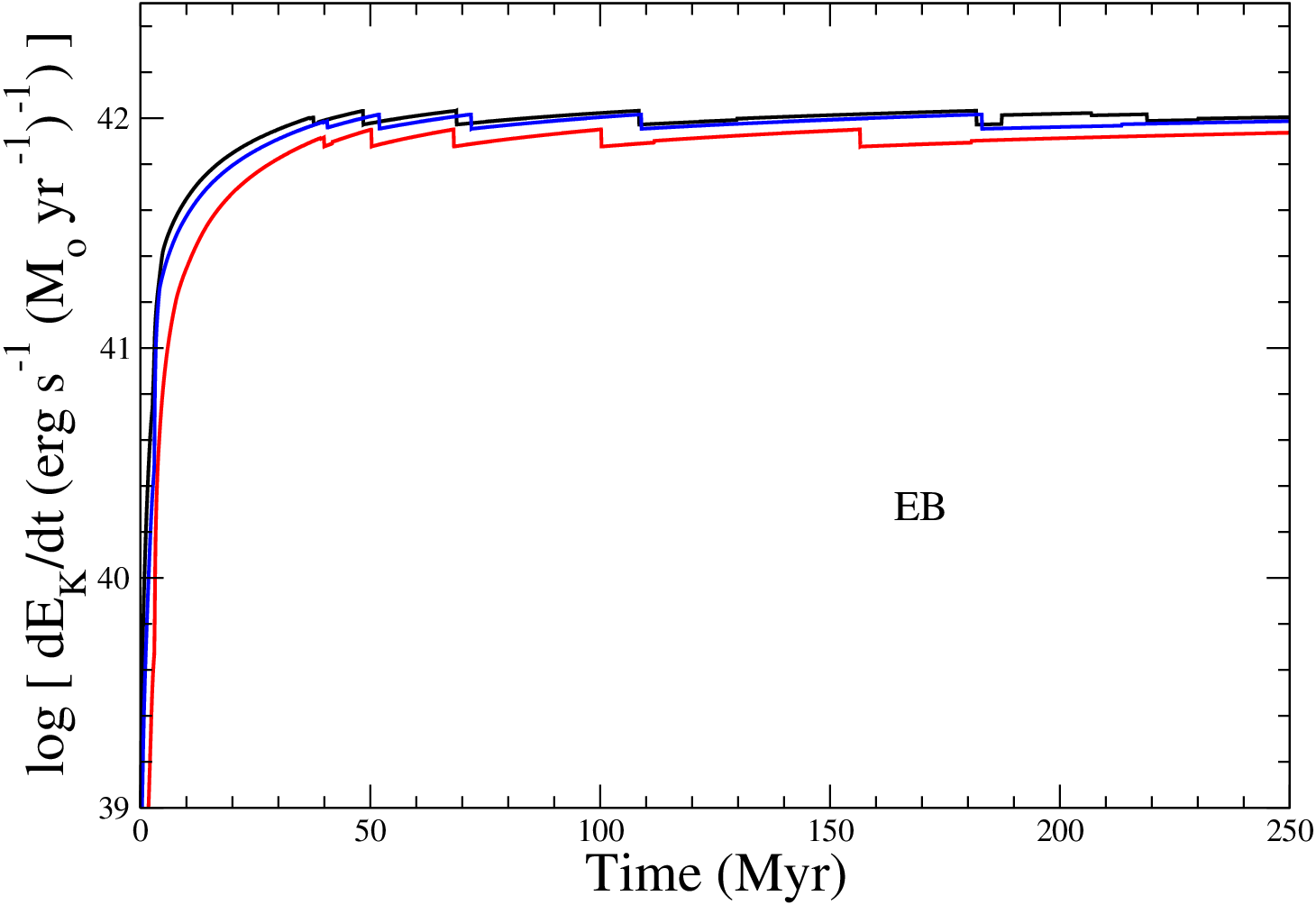}
\end{center}
\caption{Evolution of \emech\ predicted by CMHK02 for IB models (left, normalized to \sfs\ $= 1$ \msun) and by SB99 for EB models (right, normalized to \sfr\ $= 1$ \msun\ yr$^{-1}$) as a function of metallicity. From top to bottom: $Z=0.020\mbox{-black, }0.008\mbox{-blue, }0.001\mbox{-red}$.} 
\label{figemech}
\end{figure*}

\subsection{Ultraviolet and U-band continuum}

UV emission is dominated by the most massive stars in the burst (via stellar and nebular continuum radiation), and therefore constitutes a direct tracer of the presence of young, massive stars, since it is free of contamination by the continuum emission of low-mass stars. Nevertheless, interstellar extinction is very strong in this energy range, and therefore its correction plays a key role in the use of the UV continuum luminosity as a tracer of star formation. In Fig.~\ref{figuvu} we show the evolution of the luminosity at $1500$, $2000$ and $3500$ \AA\ for IB models as predicted by CMHK02,  and for EB models by SB99, for $Z=0.020$ and $Z=0.008$. No extinction has been applied in these figures. 

In IB models, the UV continuum follows the well-known trend, with a peak of the emission at $\sim3$ Myr, and a severe decrease when the most massive stars begin to explode.  The slower evolution at low metallicity is also evident in the figures. In the case of EB models, the UV continuum asymptotically stabilizes after the first $\sim50$ Myr of evolution. The stabilization is slower for \lgu, since it is contributed by stars of a lower mass than \lcuva.

We show also in Fig.~\ref{figuvu} the \lcuva\ values by  \citet{Kennicutt98} and \citet{Salim07}, after they are adapted to our IMF mass limits. Both calibrations are close to our predictions, with a perfect match with the one by \citet{Kennicutt98}  at ages around 100 Myr for solar metallicity models.  The predictions by \citet{Salim07} are somewhat overestimated with respect to both the predictions by \citet{Kennicutt98} and by us. But these authors used the \citet{Bruzual03} synthesis models in order to fit the photometric data points of their sample, assuming an exponentially-declining continuous star formation history, over which random-in-time starbursts were superimposed. The star formation histories are so different between these models that a direct comparison is not straightforward. 

\citet{Kennicutt98} argued that his calibration is applicable to bursts which have been forming stars for $100$ Myr or longer. Our results show that these calibrations can indeed be used accurately for ages above $\sim30$ Myr, since the stronger source of error will be the correction of the interstellar extinction. 

In Tables~\ref{estimatorsIBpop3} and \ref{estimatorsEBpop3} we list the corresponding \lcuva\ predictions for starbursts dominated by Pop. III stars.  As discussed above, no extinction has been assumed in this case, since no dust particles are expected to be present.

\subsection{B-, V- and K-band continuum}

In the optical and IR bands the contribution to the continuum emission by low-mass stars becomes very important as the starburst evolves. In Fig.~\ref{figbv} we show the evolution of both \lgb\ and \lgv\ for different metallicities. In IB models the formation of Wolf-Rayet stars induces some bumps between 3 and 6 Myr, due to their short lifetime and their relatively low number. These bumps are indeed a stochastic effect and should not be considered as real features. The effect of low-mass stars accumulation is clearly seen in the curves for EB models. The slope is higher than for the continuum at shorter wavelengths and, indeed, \lgb\ and \lgv\ do not stabilize even after 250 Myr of evolution. Moreover, the continuum in the optical range can be strongly contaminated by an underlying older stellar population, not related to the present episode of star formation, which makes the estimators based on \lgb\ and \lgv\ very uncertain. 

The evolution of \lgk\ is strongly affected by the formation and evolution of red supergiant stars (RSG), which are very dependent on metallicity. We plot in Fig.~\ref{figk} the evolution of \lgk\ as predicted by CMHK02 for IB models and by SB99 for EB models for three values of the metallicity.  The formation of RSG is favoured in high metallicity environments, dominating completely \lgk\ in $Z=0.020$ IB models in the period of $5 - 15$ Myr. On the other hand, their contribution for $Z=0.001$ becomes negligible (see \citet{Cervino94} for a more detailed discussion). 

In extended episodes the population of RSG stars tends to stabilize  at high metallicities. Nevertheless, the accumulation of low-mass stars, whose contribution to the K-band is important, drives a continuous increase of \lgk, which is specially evident for low metallicities where the contribution of RSG is negligible. In Fig.~\ref{figkEB} we zoom on the first 20 Myr of evolution, to show the effect of metallicity on \lgk\ early evolution.

\subsection{X-ray luminosity (\lx)}
\label{conclulx}

At the early stages of the burst, the injection of mechanical energy into the ISM is dominated by stellar winds, which tend to be stronger as the most massive stars evolve. Around $3$ Myr after the onset of the burst, massive stars end their lifetimes, and this component begins to decrease monotonically. Then the first SN start to explode, becoming the dominant sources of mechanical energy release in IB models after $3 - 4$ Myr. In EB models both components tend to be similar once the stellar population has attained a steady state, around $30 - 40$ Myr after the onset of the burst. This can be seen in Fig.~\ref{figemech}, where we have plotted the evolution of \emech\  as predicted by SB99 and CMHK02 for EB and IB models respectively, and for different metallicities. The strongest release of mechanical energy in IB models occurs at $\sim4$ Myr, when the first SN appear and add their contribution to the winds dominated by the Wolf-Rayet stars. The release of mechanical energy stabilizes in IB models after $\sim6$ Myr of evolution, since the supernova rate decreases very slowy for the next 20 Myr for a Salpeter IMF (see \citet{Cervino94}). In EB models the evolution of \emech\ stabilizes completely $30 - 40$ Myr after the onset of the burst, depending on metallicity, when both the population of the most massive stars, responsible for the winds and SN, and medium-mass stars ($M > 6 - 8$ \msun),  originator of SN as well, have reached an equilibrium.

\emech\ is very sensitive to metallicity as long as it is dominated by stellar winds, since the mass-loss rate varies largely with the metal abundance of stellar atmospheres. This is especially evident in the first 3 Myr of evolution of IB models, as shown in Fig.~\ref{figemech}, but the effect can also be seen on the lowest metallicity EB models.

As discussed above we have assumed \xeff $= 0.05$ to calibrate \lsx,  i.e., we consider that 5\% of the mechanical power that is released is reconverted into soft X-ray emission. We want to stress that while \xeff\ seems to be in the range \xeff $= 0.01 - 0.1$ for star-forming galaxies, it could attain higher or lower values in individual objects, depending on the details of the interaction between the accelerated winds and the surrounding interstellar medium. In Table \ref{estimatorsIBall} we list only the $SFS($\lsx$)$ calibration for an IB model at 5 Myr. As discussed by \citet{MasHesse08} there are large uncertainties in \lsx\ at these ages, therefore the value listed should be considered as the best average for the $4 - 6$ Myr period. The original \sfr\ calibration by \citet{MasHesse08} did not provide a value for a burst age of $250$ Myr, but as we have explained, a steady state is expected at $30 - 40$ Myr in EB models, so we assume the same estimator value for $250$ Myr as for $30$ Myr, as listed in Table \ref{estimatorsEBall}.

\section{Conclusions}

We have computed a self-consistent calibration of the most usual star formation rate tracers using evolutionary synthesis models optimized for both short-lived and extended star formation processes. We have taken into account the effects of the evolutionary state, the star formation regime and the effect of interstellar extinction on the different estimators. We have also included the prediction for population III starbursts. The main results can be summarized as follows: 

\begin{itemize}

\item The intensity of star formation episodes should be measured as a {\em star formation rate} (\sfr), in units of \msun\ yr$^{-1}$, when star formation proceeds at nearly constant rates for long periods of time (tens to hundreds of Myr), and as a {\em star formation strength} (\sfs), in units of \msun, when the episode is nearly instantaneous or short-lived (few Myr). Deriving an \sfr\ value for an instantaneous burst would not have any physical meaning. Moreover, the evolution is so rapid in these conditions that the results would be misleading and even erroneous. 

\item We have provided calibrations for \sfs\ estimators at the ages of  $4 - 6$ Myr. The ionizing power of these bursts decays so rapidly that they would show just very weak emission lines at older ages, and would not be easily recognized as massive starbursts.

\item The evolutionary state should also be taken into account even for extended episodes of star formation, whenever possible. In general most of the \sfr\ tracers have stabilized after 100~Myr of evolution at a constant rate. 

\item Some of the most used \sfr\ and \sfs\ tracers are strongly affected by interstellar extinction. We have provided their calibration as a function of \ebv\ assuming the Galactic extinction law by \citet{Cardelli89}, to allow for a more precise result if the extinction can be somehow estimated. \sfrfir\ has also been computed as a function of the extinction. For \ebv\ values above 0.5  \lfir\ can be considered as a good estimate of the bolometric luminosity of the burst. 

\item   \sfr\ and \sfs\ tracers allow to derive the amount of gas transformed into stars, but they are calibrated for a given initial mass function. The choice of different usual IMF mass limits within the range $0.1 - 120$ \msun, and IMF slopes $\alpha = 1, 3$ can lead to variations in the derived rates of up to a factor 4. A comparison of results by different authors is therefore meaningful only after a normalization to a common IMF.

\item Our calibration of star formation estimators at different energy ranges (from X-rays to radio) can also be used inversely, (taking care of the different time constants and properties of the different tracers): for example, once the \sfr\ or \sfs\ is estimated from the UV or optical continuum luminosity, the expected emission at other ranges or emission lines can be directly computed, corrected for the effects of interstellar extinction in a consistent way. Nevertheless, the inverse use of the \sfr\ and \sfs\ calibrators to derive the expected luminosity at other energy ranges has to be taken with caution, since they would predict only the luminosity associated to the star formation episode, which could be just a fraction of the total luminosity of the galaxy at certain wavelength ranges dominated by an old underlying stellar population.

\end{itemize}

\begin{acknowledgements}

JMMH and HOF are partially funded by Spanish MICINN grants CSD2006-00070 (CONSOLIDER GTC), AYA 2007-67965 and AYA2008-03467/ESP. OHF is funded by Spanish FPI grant BES-2006-13489. We thank members from the {\em Estallidos} team for giving us the idea of creating the website and for comments on its first versions, and Ra\'{u}l Guti\'{e}rrez S\'{a}nchez and Carlos Rodrigo Blanco for their help in its implementation. We also thank Miguel Cervi\~{n}o for his helpful comments about the {\em CMHK02} models. We want to acknowledge the use of the {\em Starburst99} models. We are also very grateful to D. Schaerer for providing his calculations for population III models. 

\end{acknowledgements}

\clearpage

\begin{table*}
\caption{\sfs\ estimators which do depend on \ebv\ for an IB model at age=$4$ Myr.}
\label{estimatorsIB04}
\centering
\begin{tabular}{c | c c c c c c c c c c}
\hline\hline\\
\ebv\ & \lulyalp$^{\mathrm{a}}$ & \lha$^{\mathrm{a}}$ & \lcuva$^{\mathrm{b}}$ & \lcuvb$^{\mathrm{b}}$ & \lgu$^{\mathrm{b}}$ & \lgb$^{\mathrm{b}}$ & \lgv$^{\mathrm{b}}$ & \lgk$^{\mathrm{b}}$ & \lfir$^{\mathrm{a}}$ \\
\hline\\
$0.0$ & $5.7\times 10^{-36}$ & $4.9\times 10^{-35}$ & $3.3\times 10^{-34}$ & $6.7\times 10^{-34}$ & $2.8\times 10^{-33}$ & $4.2\times 10^{-33}$ & $8.3\times 10^{-33}$ & $7.8\times 10^{-31}$ & - \\
$0.1$ & $1.5\times 10^{-35}$ & $6.2\times 10^{-35}$ & $7.0\times 10^{-34}$ & $1.5\times 10^{-33}$ & $4.4\times 10^{-33}$ & $6.1\times 10^{-33}$ & $1.1\times 10^{-32}$ & $8.1\times 10^{-31}$ & $3.0\times 10^{-37}$ \\
$0.2$ & $3.8\times 10^{-35}$ & $7.8\times 10^{-35}$ & $1.5\times 10^{-33}$ & $3.4\times 10^{-33}$ & $7.0\times 10^{-33}$ & $8.9\times 10^{-33}$ & $1.5\times 10^{-32}$ & $8.3\times 10^{-31}$ & $2.1\times 10^{-37}$ \\
$0.3$ & $9.9\times 10^{-35}$ & $9.9\times 10^{-35}$ & $3.2\times 10^{-33}$ & $7.7\times 10^{-33}$ & $1.1\times 10^{-32}$ & $1.3\times 10^{-32}$ & $2.0\times 10^{-32}$ & $8.6\times 10^{-31}$ & $1.9\times 10^{-37}$ \\
$0.4$ & $2.6\times 10^{-34}$ & $1.2\times 10^{-34}$ & $6.9\times 10^{-33}$ & $1.7\times 10^{-32}$ & $1.7\times 10^{-32}$ & $1.9\times 10^{-32}$ & $2.6\times 10^{-32}$ & $8.9\times 10^{-31}$ & $1.8\times 10^{-37}$ \\
$0.5$ & $6.6\times 10^{-34}$ & $1.6\times 10^{-34}$ & $1.5\times 10^{-32}$ & $3.9\times 10^{-32}$ & $2.7\times 10^{-32}$ & $2.8\times 10^{-32}$ & $3.5\times 10^{-32}$ & $9.2\times 10^{-31}$ & $1.7\times 10^{-37}$ \\
$0.6$ & $1.7\times 10^{-33}$ & $2.0\times 10^{-34}$ & $3.1\times 10^{-32}$ & $8.8\times 10^{-32}$ & $4.3\times 10^{-32}$ & $4.1\times 10^{-32}$ & $4.6\times 10^{-32}$ & $9.5\times 10^{-31}$ & $1.7\times 10^{-37}$ \\
$0.7$ & $4.5\times 10^{-33}$ & $2.5\times 10^{-34}$ & $6.7\times 10^{-32}$ & $2.0\times 10^{-31}$ & $6.8\times 10^{-32}$ & $5.9\times 10^{-32}$ & $6.1\times 10^{-32}$ & $9.8\times 10^{-31}$ & $1.7\times 10^{-37}$ \\
$0.8$ & $1.2\times 10^{-32}$ & $3.2\times 10^{-34}$ & $1.4\times 10^{-31}$ & $4.4\times 10^{-31}$ & $1.1\times 10^{-31}$ & $8.6\times 10^{-32}$ & $8.1\times 10^{-32}$ & $1.0\times 10^{-30}$ & $1.7\times 10^{-37}$ \\
$0.9$ & $3.0\times 10^{-32}$ & $4.0\times 10^{-34}$ & $3.1\times 10^{-31}$ & $1.0\times 10^{-30}$ & $1.7\times 10^{-31}$ & $1.3\times 10^{-31}$ & $1.1\times 10^{-31}$ & $1.0\times 10^{-30}$ & $1.7\times 10^{-37}$ \\
$1.0$ & $7.8\times 10^{-32}$ & $5.1\times 10^{-34}$ & $6.6\times 10^{-31}$ & $2.3\times 10^{-30}$ & $2.7\times 10^{-31}$ & $1.8\times 10^{-31}$ & $1.4\times 10^{-31}$ & $1.1\times 10^{-30}$ & $1.7\times 10^{-37}$ \\
\hline
\end{tabular}
\begin{list}{}{}
\item[$^{\mathrm{a}}$] Measured in \ergs\ \msun$^{-1}$.
\item[$^{\mathrm{b}}$] Measured in \ergs\ \AA$^{-1}$ \msun$^{-1}$.
\end{list}
\end{table*}

\begin{table*}
\caption{\sfs\ estimators which do depend on \ebv\ for an IB model at age=$5$ Myr.}
\label{estimatorsIB05}
\centering
\begin{tabular}{c | c c c c c c c c c c}
\hline\hline\\
\ebv\ & \lulyalp$^{\mathrm{a}}$ & \lha$^{\mathrm{a}}$ & \lcuva$^{\mathrm{b}}$ & \lcuvb$^{\mathrm{b}}$ & \lgu$^{\mathrm{b}}$ & \lgb$^{\mathrm{b}}$ & \lgv$^{\mathrm{b}}$ & \lgk$^{\mathrm{b}}$ & \lfir$^{\mathrm{a}}$ \\
\hline\\
$0.0$ & $8.6\times 10^{-36}$ & $7.4\times 10^{-35}$ & $5.0\times 10^{-34}$ & $1.1\times 10^{-33}$ & $4.9\times 10^{-33}$ & $6.1\times 10^{-33}$ & $9.0\times 10^{-33}$ & $3.4\times 10^{-31}$ & - \\
$0.1$ & $2.2\times 10^{-35}$ & $9.4\times 10^{-35}$ & $1.1\times 10^{-33}$ & $2.4\times 10^{-33}$ & $7.7\times 10^{-33}$ & $8.9\times 10^{-33}$ & $1.2\times 10^{-32}$ & $3.5\times 10^{-31}$ & $4.6\times 10^{-37}$ \\
$0.2$ & $5.8\times 10^{-35}$ & $1.2\times 10^{-34}$ & $2.3\times 10^{-33}$ & $5.4\times 10^{-33}$ & $1.2\times 10^{-32}$ & $1.3\times 10^{-32}$ & $1.6\times 10^{-32}$ & $3.6\times 10^{-31}$ & $3.3\times 10^{-37}$ \\
$0.3$ & $1.5\times 10^{-34}$ & $1.5\times 10^{-34}$ & $4.9\times 10^{-33}$ & $1.2\times 10^{-32}$ & $1.9\times 10^{-32}$ & $1.9\times 10^{-32}$ & $2.1\times 10^{-32}$ & $3.7\times 10^{-31}$ & $2.9\times 10^{-37}$ \\
$0.4$ & $3.9\times 10^{-34}$ & $1.9\times 10^{-34}$ & $1.0\times 10^{-32}$ & $2.7\times 10^{-32}$ & $3.0\times 10^{-32}$ & $2.8\times 10^{-32}$ & $2.8\times 10^{-32}$ & $3.8\times 10^{-31}$ & $2.7\times 10^{-37}$ \\
$0.5$ & $1.0\times 10^{-33}$ & $2.4\times 10^{-34}$ & $2.2\times 10^{-32}$ & $6.1\times 10^{-32}$ & $4.7\times 10^{-32}$ & $4.0\times 10^{-32}$ & $3.7\times 10^{-32}$ & $4.0\times 10^{-31}$ & $2.6\times 10^{-37}$ \\
$0.6$ & $2.6\times 10^{-33}$ & $3.0\times 10^{-34}$ & $4.8\times 10^{-32}$ & $1.4\times 10^{-31}$ & $7.5\times 10^{-32}$ & $5.9\times 10^{-32}$ & $5.0\times 10^{-32}$ & $4.1\times 10^{-31}$ & $2.6\times 10^{-37}$ \\
$0.7$ & $6.7\times 10^{-33}$ & $3.8\times 10^{-34}$ & $1.0\times 10^{-31}$ & $3.1\times 10^{-31}$ & $1.2\times 10^{-31}$ & $8.6\times 10^{-32}$ & $6.6\times 10^{-32}$ & $4.2\times 10^{-31}$ & $2.6\times 10^{-37}$ \\
$0.8$ & $1.7\times 10^{-32}$ & $4.8\times 10^{-34}$ & $2.2\times 10^{-31}$ & $7.0\times 10^{-31}$ & $1.9\times 10^{-31}$ & $1.3\times 10^{-31}$ & $8.8\times 10^{-32}$ & $4.4\times 10^{-31}$ & $2.6\times 10^{-37}$ \\
$0.9$ & $4.5\times 10^{-32}$ & $6.1\times 10^{-34}$ & $4.7\times 10^{-31}$ & $1.6\times 10^{-30}$ & $2.9\times 10^{-31}$ & $1.8\times 10^{-31}$ & $1.2\times 10^{-31}$ & $4.5\times 10^{-31}$ & $2.6\times 10^{-37}$ \\
$1.0$ & $1.2\times 10^{-31}$ & $7.7\times 10^{-34}$ & $1.0\times 10^{-30}$ & $3.5\times 10^{-30}$ & $4.6\times 10^{-31}$ & $2.7\times 10^{-31}$ & $1.6\times 10^{-31}$ & $4.7\times 10^{-31}$ & $2.5\times 10^{-37}$ \\
\hline
\end{tabular}
\begin{list}{}{}
\item[$^{\mathrm{a}}$] Measured in \ergs\ \msun$^{-1}$.
\item[$^{\mathrm{b}}$] Measured in \ergs\ \AA$^{-1}$ \msun$^{-1}$.
\end{list}
\end{table*}

\begin{table*}
\caption{\sfs\ estimators which do depend on \ebv\ for an IB model at age=$6$ Myr.}
\label{estimatorsIB06}
\centering
\begin{tabular}{c | c c c c c c c c c c}
\hline\hline\\
\ebv\ & \lulyalp$^{\mathrm{a}}$ & \lha$^{\mathrm{a}}$ & \lcuva$^{\mathrm{b}}$ & \lcuvb$^{\mathrm{b}}$ & \lgu$^{\mathrm{b}}$ & \lgb$^{\mathrm{b}}$ & \lgv$^{\mathrm{b}}$ & \lgk$^{\mathrm{b}}$ & \lfir$^{\mathrm{a}}$ \\
\hline\\
$0.0$ & $3.8\times 10^{-35}$ & $3.3\times 10^{-34}$ & $6.9\times 10^{-34}$ & $1.5\times 10^{-33}$ & $7.0\times 10^{-33}$ & $5.8\times 10^{-33}$ & $8.6\times 10^{-33}$ & $2.5\times 10^{-31}$ & - \\
$0.1$ & $9.9\times 10^{-35}$ & $4.2\times 10^{-34}$ & $1.5\times 10^{-33}$ & $3.3\times 10^{-33}$ & $1.1\times 10^{-32}$ & $8.4\times 10^{-33}$ & $1.1\times 10^{-32}$ & $2.6\times 10^{-31}$ & $6.7\times 10^{-37}$ \\
$0.2$ & $2.6\times 10^{-34}$ & $5.3\times 10^{-34}$ & $3.2\times 10^{-33}$ & $7.4\times 10^{-33}$ & $1.7\times 10^{-32}$ & $1.2\times 10^{-32}$ & $1.5\times 10^{-32}$ & $2.7\times 10^{-31}$ & $4.6\times 10^{-37}$ \\
$0.3$ & $6.7\times 10^{-34}$ & $6.7\times 10^{-34}$ & $6.7\times 10^{-33}$ & $1.7\times 10^{-32}$ & $2.7\times 10^{-32}$ & $1.8\times 10^{-32}$ & $2.0\times 10^{-32}$ & $2.7\times 10^{-31}$ & $4.0\times 10^{-37}$ \\
$0.4$ & $1.7\times 10^{-33}$ & $8.4\times 10^{-34}$ & $1.4\times 10^{-32}$ & $3.8\times 10^{-32}$ & $4.3\times 10^{-32}$ & $2.6\times 10^{-32}$ & $2.7\times 10^{-32}$ & $2.8\times 10^{-31}$ & $3.8\times 10^{-37}$ \\
$0.5$ & $4.5\times 10^{-33}$ & $1.1\times 10^{-33}$ & $3.1\times 10^{-32}$ & $8.5\times 10^{-32}$ & $6.8\times 10^{-32}$ & $3.8\times 10^{-32}$ & $3.6\times 10^{-32}$ & $2.9\times 10^{-31}$ & $3.7\times 10^{-37}$ \\
$0.6$ & $1.2\times 10^{-32}$ & $1.3\times 10^{-33}$ & $6.6\times 10^{-32}$ & $1.9\times 10^{-31}$ & $1.1\times 10^{-31}$ & $5.6\times 10^{-32}$ & $4.8\times 10^{-32}$ & $3.0\times 10^{-31}$ & $3.6\times 10^{-37}$ \\
$0.7$ & $3.0\times 10^{-32}$ & $1.7\times 10^{-33}$ & $1.4\times 10^{-31}$ & $4.3\times 10^{-31}$ & $1.7\times 10^{-31}$ & $8.2\times 10^{-32}$ & $6.3\times 10^{-32}$ & $3.1\times 10^{-31}$ & $3.5\times 10^{-37}$ \\
$0.8$ & $7.8\times 10^{-32}$ & $2.1\times 10^{-33}$ & $3.0\times 10^{-31}$ & $9.7\times 10^{-31}$ & $2.6\times 10^{-31}$ & $1.2\times 10^{-31}$ & $8.4\times 10^{-32}$ & $3.2\times 10^{-31}$ & $3.5\times 10^{-37}$ \\
$0.9$ & $2.0\times 10^{-31}$ & $2.7\times 10^{-33}$ & $6.5\times 10^{-31}$ & $2.2\times 10^{-30}$ & $4.2\times 10^{-31}$ & $1.7\times 10^{-31}$ & $1.1\times 10^{-31}$ & $3.3\times 10^{-31}$ & $3.5\times 10^{-37}$ \\
$1.0$ & $5.3\times 10^{-31}$ & $3.4\times 10^{-33}$ & $1.4\times 10^{-30}$ & $4.9\times 10^{-30}$ & $6.6\times 10^{-31}$ & $2.5\times 10^{-31}$ & $1.5\times 10^{-31}$ & $3.4\times 10^{-31}$ & $3.5\times 10^{-37}$ \\
\hline
\end{tabular}
\begin{list}{}{}
\item[$^{\mathrm{a}}$] Measured in \ergs\ \msun$^{-1}$.
\item[$^{\mathrm{b}}$] Measured in \ergs\ \AA$^{-1}$ \msun$^{-1}$.
\end{list}
\end{table*}

\begin{table*}
\caption{\sfr\ estimators which do not depend on \ebv.}
\label{estimatorsEBall}
\centering
\begin{tabular}{c | c c c}
\hline\hline\\
Magnitude & EB ($10$ Myr) & EB ($30$ Myr) & EB ($250$ Myr) \\
\hline\\
\nlyc\ ($s^{-1} ($\msun\ yr$^{-1})^{-1}$) & $4.1\times 10^{-54}$ & $4.1\times 10^{-54}$ & $4.1\times 10^{-54}$ \\ 
\lradio\ (\ergs\ Hz$^{-1}$ (\msun\ yr$^{-1})^{-1}$) & $9.3\times 10^{-29}$ & $4.3\times 10^{-29}$ &  $3.8\times 10^{-29}$ \\ 
\lsx\ (\ergs\ (\msun\ yr$^{-1})^{-1}$) & $8\times 10^{-41}$ & $2\times 10^{-41}$ &  $2\times 10^{-41}$ \\ 
\hline
\end{tabular}
\end{table*}

\clearpage

\begin{table*}
\caption{\sfr\ estimators which do depend on \ebv\ for an EB model at age=$10$ Myr.}
\label{estimatorsEB010}
\centering
\begin{tabular}{c | c c c c c c c c c c}
\hline\hline\\
\ebv\ & \lulyalp$^{\mathrm{a}}$ & \lha$^{\mathrm{a}}$ & \lcuva$^{\mathrm{b}}$ & \lcuvb$^{\mathrm{b}}$ & \lgu$^{\mathrm{b}}$ & \lgb$^{\mathrm{b}}$ & \lgv$^{\mathrm{b}}$ & \lgk$^{\mathrm{b}}$ & \lfir$^{\mathrm{a}}$ \\
\hline\\
$0.0$ & $3.5\times 10^{-43}$ & $3.0\times 10^{-42}$ & $4.1\times 10^{-41}$ & $8.4\times 10^{-41}$ & $3.4\times 10^{-40}$ & $6.5\times 10^{-40}$ & $1.1\times 10^{-39}$ & $2.3\times 10^{-38}$ & - \\
$0.1$ & $9.0\times 10^{-43}$ & $3.8\times 10^{-42}$ & $8.8\times 10^{-41}$ & $1.9\times 10^{-40}$ & $5.3\times 10^{-40}$ & $9.5\times 10^{-40}$ & $1.5\times 10^{-39}$ & $2.4\times 10^{-38}$ & $3.3\times 10^{-44}$ \\
$0.2$ & $2.3\times 10^{-42}$ & $4.8\times 10^{-42}$ & $1.9\times 10^{-40}$ & $4.3\times 10^{-40}$ & $8.4\times 10^{-40}$ & $1.4\times 10^{-39}$ & $1.9\times 10^{-39}$ & $2.5\times 10^{-38}$ & $2.4\times 10^{-44}$ \\
$0.3$ & $6.1\times 10^{-42}$ & $6.1\times 10^{-42}$ & $4.0\times 10^{-40}$ & $9.6\times 10^{-40}$ & $1.3\times 10^{-39}$ & $2.0\times 10^{-39}$ & $2.6\times 10^{-39}$ & $2.6\times 10^{-38}$ & $2.2\times 10^{-44}$ \\
$0.4$ & $1.6\times 10^{-41}$ & $7.7\times 10^{-42}$ & $8.6\times 10^{-40}$ & $2.2\times 10^{-39}$ & $2.1\times 10^{-39}$ & $2.9\times 10^{-39}$ & $3.4\times 10^{-39}$ & $2.6\times 10^{-38}$ & $2.0\times 10^{-44}$ \\
$0.5$ & $4.1\times 10^{-41}$ & $9.7\times 10^{-42}$ & $1.8\times 10^{-39}$ & $4.9\times 10^{-39}$ & $3.3\times 10^{-39}$ & $4.3\times 10^{-39}$ & $4.6\times 10^{-39}$ & $2.7\times 10^{-38}$ & $2.0\times 10^{-44}$ \\
$0.6$ & $1.1\times 10^{-40}$ & $1.2\times 10^{-41}$ & $4.0\times 10^{-39}$ & $1.1\times 10^{-38}$ & $5.2\times 10^{-39}$ & $6.3\times 10^{-39}$ & $6.1\times 10^{-39}$ & $2.8\times 10^{-38}$ & $2.0\times 10^{-44}$ \\
$0.7$ & $2.7\times 10^{-40}$ & $1.5\times 10^{-41}$ & $8.5\times 10^{-39}$ & $2.5\times 10^{-38}$ & $8.2\times 10^{-39}$ & $9.1\times 10^{-39}$ & $8.1\times 10^{-39}$ & $2.9\times 10^{-38}$ & $1.9\times 10^{-44}$ \\
$0.8$ & $7.1\times 10^{-40}$ & $2.0\times 10^{-41}$ & $1.8\times 10^{-38}$ & $5.6\times 10^{-38}$ & $1.3\times 10^{-38}$ & $1.3\times 10^{-38}$ & $1.1\times 10^{-38}$ & $3.0\times 10^{-38}$ & $1.9\times 10^{-44}$ \\
$0.9$ & $1.8\times 10^{-39}$ & $2.5\times 10^{-41}$ & $3.9\times 10^{-38}$ & $1.3\times 10^{-37}$ & $2.0\times 10^{-38}$ & $1.9\times 10^{-38}$ & $1.4\times 10^{-38}$ & $3.1\times 10^{-38}$ & $1.9\times 10^{-44}$ \\
$1.0$ & $4.8\times 10^{-39}$ & $3.1\times 10^{-41}$ & $8.3\times 10^{-38}$ & $2.8\times 10^{-37}$ & $3.2\times 10^{-38}$ & $2.8\times 10^{-38}$ & $1.9\times 10^{-38}$ & $3.2\times 10^{-38}$ & $1.9\times 10^{-44}$ \\
\hline
\end{tabular}
\begin{list}{}{}
\item[$^{\mathrm{a}}$] Measured in \ergs\ (\msun\ yr$^{-1})^{-1}$.
\item[$^{\mathrm{b}}$] Measured in \ergs\ \AA$^{-1}$ (\msun\ yr$^{-1})^{-1}$.
\end{list}
\end{table*}

\begin{table*}
\caption{\sfr\ estimators which do depend on \ebv\ for an EB model at age=$30$ Myr.}
\label{estimatorsEB030}
\centering
\begin{tabular}{c | c c c c c c c c c c}
\hline\hline\\
\ebv\ & \lulyalp$^{\mathrm{a}}$ & \lha$^{\mathrm{a}}$ & \lcuva$^{\mathrm{b}}$ & \lcuvb$^{\mathrm{b}}$ & \lgu$^{\mathrm{b}}$ & \lgb$^{\mathrm{b}}$ & \lgv$^{\mathrm{b}}$ & \lgk$^{\mathrm{b}}$ & \lfir$^{\mathrm{a}}$ \\
\hline\\
$0.0$ & $3.5\times 10^{-43}$ & $3.0\times 10^{-42}$ & $3.3\times 10^{-41}$ & $6.6\times 10^{-41}$ & $2.6\times 10^{-40}$ & $4.2\times 10^{-40}$ & $7.2\times 10^{-40}$ & $1.2\times 10^{-38}$ & - \\
$0.1$ & $9.0\times 10^{-43}$ & $3.8\times 10^{-42}$ & $7.1\times 10^{-41}$ & $1.5\times 10^{-40}$ & $4.1\times 10^{-40}$ & $6.1\times 10^{-40}$ & $9.6\times 10^{-40}$ & $1.3\times 10^{-38}$ & $2.7\times 10^{-44}$ \\
$0.2$ & $2.3\times 10^{-42}$ & $4.8\times 10^{-42}$ & $1.5\times 10^{-40}$ & $3.3\times 10^{-40}$ & $6.5\times 10^{-40}$ & $8.9\times 10^{-40}$ & $1.3\times 10^{-39}$ & $1.3\times 10^{-38}$ & $2.0\times 10^{-44}$ \\
$0.3$ & $6.1\times 10^{-42}$ & $6.1\times 10^{-42}$ & $3.3\times 10^{-40}$ & $7.5\times 10^{-40}$ & $1.0\times 10^{-39}$ & $1.3\times 10^{-39}$ & $1.7\times 10^{-39}$ & $1.4\times 10^{-38}$ & $1.8\times 10^{-44}$ \\
$0.4$ & $1.6\times 10^{-41}$ & $7.7\times 10^{-42}$ & $7.0\times 10^{-40}$ & $1.7\times 10^{-39}$ & $1.6\times 10^{-39}$ & $1.9\times 10^{-39}$ & $2.2\times 10^{-39}$ & $1.4\times 10^{-38}$ & $1.7\times 10^{-44}$ \\
$0.5$ & $4.1\times 10^{-41}$ & $9.7\times 10^{-42}$ & $1.5\times 10^{-39}$ & $3.8\times 10^{-39}$ & $2.5\times 10^{-39}$ & $2.8\times 10^{-39}$ & $3.0\times 10^{-39}$ & $1.5\times 10^{-38}$ & $1.6\times 10^{-44}$ \\
$0.6$ & $1.1\times 10^{-40}$ & $1.2\times 10^{-41}$ & $3.2\times 10^{-39}$ & $8.6\times 10^{-39}$ & $4.0\times 10^{-39}$ & $4.1\times 10^{-39}$ & $4.0\times 10^{-39}$ & $1.5\times 10^{-38}$ & $1.6\times 10^{-44}$ \\
$0.7$ & $2.7\times 10^{-40}$ & $1.5\times 10^{-41}$ & $6.8\times 10^{-39}$ & $1.9\times 10^{-38}$ & $6.3\times 10^{-39}$ & $5.9\times 10^{-39}$ & $5.3\times 10^{-39}$ & $1.6\times 10^{-38}$ & $1.6\times 10^{-44}$ \\
$0.8$ & $7.1\times 10^{-40}$ & $1.9\times 10^{-41}$ & $1.5\times 10^{-38}$ & $4.4\times 10^{-38}$ & $9.9\times 10^{-39}$ & $8.6\times 10^{-39}$ & $7.0\times 10^{-39}$ & $1.6\times 10^{-38}$ & $1.6\times 10^{-44}$ \\
$0.9$ & $1.8\times 10^{-39}$ & $2.5\times 10^{-41}$ & $3.1\times 10^{-38}$ & $9.8\times 10^{-38}$ & $1.6\times 10^{-38}$ & $1.3\times 10^{-38}$ & $9.4\times 10^{-39}$ & $1.7\times 10^{-38}$ & $1.6\times 10^{-44}$ \\
$1.0$ & $4.8\times 10^{-39}$ & $3.1\times 10^{-41}$ & $6.7\times 10^{-38}$ & $2.2\times 10^{-37}$ & $2.4\times 10^{-38}$ & $1.8\times 10^{-38}$ & $1.2\times 10^{-38}$ & $1.7\times 10^{-38}$ & $1.5\times 10^{-44}$ \\
\hline
\end{tabular}
\begin{list}{}{}
\item[$^{\mathrm{a}}$] Measured in \ergs\ (\msun\ yr$^{-1})^{-1}$.
\item[$^{\mathrm{b}}$] Measured in \ergs\ \AA$^{-1}$ (\msun\ yr$^{-1})^{-1}$.
\end{list}
\end{table*}

\begin{table*}
\caption{\sfr\ estimators which do depend on \ebv\ for an EB model at age=$250$ Myr.}
\label{estimatorsEB250}
\centering
\begin{tabular}{c | c c c c c c c c c c}
\hline\hline\\
\ebv\ & \lulyalp$^{\mathrm{a}}$ & \lha$^{\mathrm{a}}$ & \lcuva$^{\mathrm{b}}$ & \lcuvb$^{\mathrm{b}}$ & \lgu$^{\mathrm{b}}$ & \lgb$^{\mathrm{b}}$ & \lgv$^{\mathrm{b}}$ & \lgk$^{\mathrm{b}}$ & \lfir$^{\mathrm{a}}$ \\
\hline\\
$0.0$ & $3.5\times 10^{-43}$ & $3.0\times 10^{-42}$ & $2.7\times 10^{-41}$ & $5.1\times 10^{-41}$ & $1.8\times 10^{-40}$ & $2.2\times 10^{-40}$ & $3.6\times 10^{-40}$ & $6.7\times 10^{-39}$ & - \\
$0.1$ & $9.0\times 10^{-43}$ & $3.8\times 10^{-42}$ & $5.8\times 10^{-41}$ & $1.2\times 10^{-40}$ & $2.9\times 10^{-40}$ & $3.2\times 10^{-40}$ & $4.8\times 10^{-40}$ & $6.9\times 10^{-39}$ & $2.3\times 10^{-44}$ \\
$0.2$ & $2.3\times 10^{-42}$ & $4.8\times 10^{-42}$ & $1.2\times 10^{-40}$ & $2.6\times 10^{-40}$ & $4.5\times 10^{-40}$ & $4.7\times 10^{-40}$ & $6.4\times 10^{-40}$ & $7.1\times 10^{-39}$ & $1.6\times 10^{-44}$ \\
$0.3$ & $6.1\times 10^{-42}$ & $6.1\times 10^{-42}$ & $2.6\times 10^{-40}$ & $5.8\times 10^{-40}$ & $7.1\times 10^{-40}$ & $6.8\times 10^{-40}$ & $8.6\times 10^{-40}$ & $7.4\times 10^{-39}$ & $1.4\times 10^{-44}$ \\
$0.4$ & $1.6\times 10^{-41}$ & $7.7\times 10^{-42}$ & $5.6\times 10^{-40}$ & $1.3\times 10^{-39}$ & $1.1\times 10^{-39}$ & $1.0\times 10^{-39}$ & $1.1\times 10^{-39}$ & $7.6\times 10^{-39}$ & $1.3\times 10^{-44}$ \\
$0.5$ & $4.1\times 10^{-41}$ & $9.7\times 10^{-42}$ & $1.2\times 10^{-39}$ & $3.0\times 10^{-39}$ & $1.8\times 10^{-39}$ & $1.5\times 10^{-39}$ & $1.5\times 10^{-39}$ & $7.9\times 10^{-39}$ & $1.3\times 10^{-44}$ \\
$0.6$ & $1.1\times 10^{-40}$ & $1.2\times 10^{-41}$ & $2.6\times 10^{-39}$ & $6.7\times 10^{-39}$ & $2.8\times 10^{-39}$ & $2.1\times 10^{-39}$ & $2.0\times 10^{-39}$ & $8.1\times 10^{-39}$ & $1.3\times 10^{-44}$ \\
$0.7$ & $2.7\times 10^{-40}$ & $1.5\times 10^{-41}$ & $5.5\times 10^{-39}$ & $1.5\times 10^{-38}$ & $4.4\times 10^{-39}$ & $3.1\times 10^{-39}$ & $2.7\times 10^{-39}$ & $8.4\times 10^{-39}$ & $1.2\times 10^{-44}$ \\
$0.8$ & $7.1\times 10^{-40}$ & $1.9\times 10^{-41}$ & $1.2\times 10^{-38}$ & $3.4\times 10^{-38}$ & $6.9\times 10^{-39}$ & $4.5\times 10^{-39}$ & $3.6\times 10^{-39}$ & $8.7\times 10^{-39}$ & $1.2\times 10^{-44}$ \\
$0.9$ & $1.8\times 10^{-39}$ & $2.5\times 10^{-41}$ & $2.5\times 10^{-38}$ & $7.6\times 10^{-38}$ & $1.1\times 10^{-38}$ & $6.6\times 10^{-39}$ & $4.7\times 10^{-39}$ & $8.9\times 10^{-39}$ & $1.2\times 10^{-44}$ \\
$1.0$ & $4.8\times 10^{-39}$ & $3.1\times 10^{-41}$ & $5.4\times 10^{-38}$ & $1.7\times 10^{-37}$ & $1.7\times 10^{-38}$ & $9.7\times 10^{-39}$ & $6.3\times 10^{-39}$ & $9.2\times 10^{-39}$ & $1.2\times 10^{-44}$ \\
\hline
\end{tabular}
\begin{list}{}{}
\item[$^{\mathrm{a}}$] Measured in \ergs\ (\msun\ yr$^{-1})^{-1}$.
\item[$^{\mathrm{b}}$] Measured in \ergs\ \AA$^{-1}$ (\msun\ yr$^{-1})^{-1}$.
\end{list}
\end{table*}

\begin{table*}
\centering
\caption{\sfs\ estimators based on \nlyc\ and \lcuva\ for Pop. III.}
\label{estimatorsIBpop3}
\begin{tabular}{c|cc|cc|cc}
\hline\hline &&&&&&\\
 & \multicolumn{2}{c|}{IB($2-120$ \msun) $^{\mathrm{\dag}}$} & \multicolumn{2}{c|}{IB($1-100$ \msun)} &  \multicolumn{2}{c}{IB($1-500$ \msun)} \\
\hline
Age & \nlyc$^{\mathrm{a}}$ & \lcuva$^{\mathrm{b}}$ & \nlyc$^{\mathrm{a}}$ & \lcuva$^{\mathrm{b}}$ & \nlyc$^{\mathrm{a}}$ & \lcuva$^{\mathrm{b}}$ \\
\hline &&&&&&\\
$4$ Myr & $1.45\times 10^{-47}$ & $8.30\times 10^{-34}$ & $1.94\times 10^{-47}$ & $1.11\times 10^{-33}$ & $2.15\times 10^{-47}$ & $1.23\times 10^{-33}$ \\
$5$ Myr & $1.95\times 10^{-47}$ & $9.98\times 10^{-34}$ & $2.61\times 10^{-47}$ & $1.34\times 10^{-33}$ & $2.89\times 10^{-47}$ & $1.48\times 10^{-33}$ \\
$6$ Myr & $2.53\times 10^{-47}$ & $1.14\times 10^{-33}$ & $3.40\times 10^{-47}$ & $1.52\times 10^{-33}$ & $3.76\times 10^{-47}$ & $1.69\times 10^{-33}$ \\
\hline
\end{tabular}
\begin{list}{}{}
\item[$^{\mathrm{\dag}}$] Values for $2-120$ \msun\ were obtained analytically from those for $1-100$ \msun.
\item[$^{\mathrm{a}}$] Measured in $s^{-1}$ \msun$^{-1}$.
\item[$^{\mathrm{b}}$] Measured in \ergs\ \AA$^{-1}$ \msun$^{-1}$.
\end{list}
\end{table*}

\begin{table*}
\centering
\caption{\sfr\ estimators based on \nlyc\ and \lcuva\ for Pop. III.}
\label{estimatorsEBpop3}
\begin{tabular}{c|cc|cc|cc}
\hline\hline &&&&&&\\
 & \multicolumn{2}{c|}{EB($2-120$ \msun) $^{\mathrm{\dag}}$} & \multicolumn{2}{c|}{EB($1-100$ \msun)} &  \multicolumn{2}{c}{EB($1-500$ \msun)}  \\
\hline
Age & \nlyc$^{\mathrm{a}}$ & \lcuva$^{\mathrm{b}}$ & \nlyc$^{\mathrm{a}}$ & \lcuva$^{\mathrm{b}}$ & \nlyc$^{\mathrm{a}}$ & \lcuva$^{\mathrm{b}}$ \\
\hline &&&&&&\\ 
$10$ Myr & $1.27\times 10^{-54}$ & $7.96\times 10^{-41}$ & $1.70\times 10^{-54}$ & $1.07\times 10^{-40}$ & $1.26\times 10^{-54}$ & $8.67\times 10^{-41}$ \\
$30$ Myr & $1.15\times 10^{-54}$ & $5.60\times 10^{-41}$ & $1.55\times 10^{-54}$ & $7.50\times 10^{-41}$ & $1.18\times 10^{-54}$ & $6.61\times 10^{-41}$ \\
$250$ Myr & $1.14\times 10^{-54}$ & $3.53\times 10^{-41}$ & $1.53\times 10^{-54}$ & $4.73\times 10^{-41}$ & $1.17\times 10^{-54}$ & $4.52\times 10^{-41}$ \\
\hline
\end{tabular}
\begin{list}{}{}
\item[$^{\mathrm{\dag}}$] Values for $2-120$ \msun\ were obtained analytically from those for $1-100$ \msun.
\item[$^{\mathrm{a}}$] Measured in $s^{-1}$ (\msun\ yr$^{-1})^{-1}$.
\item[$^{\mathrm{b}}$] Measured in \ergs\ \AA$^{-1}$ (\msun\ yr$^{-1})^{-1}$.
\end{list}
\end{table*}

\begin{table*}
\centering
\caption{IMF correction factors for all \sfs\ estimators, to be applied on values from Tables \ref{estimatorsIBall}-\ref{estimatorsIB06} when IMF mass limits $0.1-100$ \msun\ or $1-100$ \msun\ are assumed. See Sect. \ref{imfcorr} for a more thorough discussion.}
\label{corfactorsIB}
\begin{tabular}{c|c|c}
\hline\hline &&\\
Mass limits & IB($0.1-100$ \msun) & IB($1-100$ \msun) \\
\hline
 Corr. factor & $3.41$ & $1.34$\\
\hline
\end{tabular}
\end{table*}

\begin{table*}
\centering
\caption{IMF correction factors for \sfr\ estimators, to be applied on values from Tables \ref{estimatorsEBall}-\ref{estimatorsEB250} when IMF mass limits of $0.1-100$ \msun\ or $1-100$ \msun\ are considered. See Sect. \ref{imfcorr} for a more thorough discussion.}
\label{corfactorsEB}
\begin{tabular}{c|c|c}
\hline\hline &&\\
Magnitude & EB($0.1-100$ \msun) & EB($1-100$ \msun) \\
\hline
\lfir\ & $3.54$ & $1.36$\\
\nlyc\ & $3.94$ & $1.47$\\
\lradio\ & $3.51$ & $1.37$\\
\lsx\ $^{\mathrm{\dag}}$ & $3.41$ & $1.34$\\
\lcuva\ & $3.54$ & $1.37$\\
\lcuvb\ & $3.53$ & $1.36$\\
\lgu\ & $3.47$ & $1.34$\\
\lgb\ & $3.28$ & $1.28$\\
\lgv\ & $3.26$ & $1.27$\\
\lgk\ & $3.36$ & $1.32$\\
\hline
\end{tabular}
\begin{list}{}{}
\item[$^{\mathrm{\dag}}$] The values taken for \lsx\ were calculated analytically.
\end{list}
\end{table*}

\begin{table*}
\centering
\caption{IMF correction factors for \sfs\ estimators, to be applied on values from Tables \ref{estimatorsIBall}-\ref{estimatorsIB06} when IMF slope $\alpha=1$, $3$.}
\label{corfactorsIBalpha}
\begin{tabular}{c|ccc|ccc}
\hline\hline &&&&&&\\
 & \multicolumn{3}{c|}{IB($\alpha=1$)} & \multicolumn{3}{c}{IB($\alpha=3$)} \\
\hline &&&&&&\\
Magnitude & $4$ Myr & $5$ Myr & $6$ Myr & $4$ Myr & $5$ Myr & $6$ Myr\\
\hline
\lfir & $0.61$ & $0.93$ & $1.22$ & $2.41$ & $2.03$ & $1.81$ \\
\nlyc & $0.54$ & $0.85$ & $1.17$ & $2.76$ & $2.31$ & $2.01$ \\
\lradio & $0.32$ & $0.52$ & $0.71$ & $3.69$ & $2.99$ & $2.60$ \\
\lcuva & $0.63$ & $0.97$ & $1.29$ & $2.36$ & $1.99$ & $1.76$ \\
\lcuvb & $0.61$ & $0.95$ & $1.27$ & $2.40$ & $1.99$ & $1.76$ \\
\lgu & $0.55$ & $0.90$ & $1.17$ & $2.49$ & $1.99$ & $1.78$ \\
\lgb & $0.50$ & $0.78$ & $0.95$ & $2.55$ & $2.09$ & $1.96$ \\
\lgv & $0.49$ & $0.71$ & $0.89$ & $2.59$ & $2.26$ & $2.10$ \\
\lgk & $0.48$ & $0.63$ & $0.80$ & $2.75$ & $2.64$ & $2.40$ \\
\hline
\end{tabular}
\end{table*}

\begin{table*}
\centering
\caption{IMF correction factors for \sfr\ estimators, to be applied on values from Tables \ref{estimatorsEBall}-\ref{estimatorsEB250} when IMF slope $\alpha=1$, $3$.}
\label{corfactorsEBalpha}
\begin{tabular}{c|ccc|ccc}
\hline\hline &&&&&&\\
 & \multicolumn{3}{c|}{EB($\alpha=1$)} & \multicolumn{3}{c}{EB($\alpha=3$)} \\
\hline &&&&&&\\
Magnitude & $10$ Myr & $30$ Myr & $250$ Myr & $10$ Myr & $30$ Myr & $250$ Myr\\
\hline 
\lfir & $0.37$ & $0.45$ & $0.54$ & $2.55$ & $2.09$ & $1.55$ \\
\nlyc & $0.25$ & $0.25$ & $0.25$ & $3.68$ & $3.67$ & $3.67$ \\
\lradio & $0.37$ & $0.66$ & $0.74$ & $2.95$ & $1.89$ & $1.75$ \\
\lcuva & $0.41$ & $0.49$ & $0.60$ & $2.41$ & $1.97$ & $1.49$ \\
\lcuvb & $0.40$ & $0.50$ & $0.63$ & $2.39$ & $1.92$ & $1.40$ \\
\lgu & $0.36$ & $0.46$ & $0.64$ & $2.50$ & $1.92$ & $1.26$ \\
\lgb & $0.45$ & $0.64$ & $1.16$ & $2.20$ & $1.61$ & $0.98$ \\
\lgv & $0.46$ & $0.66$ & $1.21$ & $2.27$ & $1.68$ & $1.00$ \\
\lgk & $0.60$ & $0.90$ & $1.52$ & $2.34$ & $1.82$ & $1.16$ \\
\hline
\end{tabular}
\end{table*}

\end{document}